\documentclass[journal]{IEEEtran}

\usepackage{amsmath,graphicx}
\usepackage{amssymb}
\usepackage{array}
\usepackage{slashbox}
\usepackage{amsthm}
\usepackage{multirow}
\usepackage{picinpar}
\usepackage{cases}
\usepackage[colorlinks,
            linkcolor=red,
            anchorcolor=green,
            citecolor=blue]{hyperref}
\usepackage{fourier}
\newtheorem{Theorem}{Theorem}[section]
\newtheorem{Lemma}[Theorem]{Lemma}

\newtheorem{Example}[Theorem]{Example}
\newtheorem{Remark}[Theorem]{Remark}

\newtheorem{Definition}[Theorem]{Definition}

\newcounter{claim_nb}[Theorem]
\newcommand{\wuhao}{\fontsize{10pt}{\baselineskip}\selectfont}
\newcommand{\xiaowuhao}{\fontsize{8pt}{\baselineskip}\selectfont}

\begin{document}

\title{Exact Recovery for Sparse Signal via Weighted $\ell_1$ Minimization}

\author{Shenglong Zhou, Naihua Xiu, Yingnan Wang, Lingchen Kong
\thanks{Dec 8, 2013. Department of Applied Mathematics, Beijing Jiaotong University, Beijing 100044, P. R. China (e-mail: longnan\_zsl@163.com,  nhxiu@bjtu.edu.cn, wyn1982@hotmail.com, konglchen@126.com).

Revised at Jan 28, 2014.}}

\maketitle

\begin{abstract}
Numerical experiments in literature on compressed sensing have indicated that the reweighted $\ell_1$ minimization performs exceptionally well in recovering sparse signal. In this paper, we develop exact recovery conditions and algorithm for sparse signal via weighted $\ell_1$ minimization from the insight of the classical NSP (null space property) and RIC (restricted isometry constant) bound. We first introduce the concept of WNSP (weighted null space property) and reveal that it is a necessary and sufficient condition for exact recovery. We then prove that the RIC bound by weighted $\ell_1$ minimization is
$$\delta_{ak}< \sqrt{\frac{a-1}{a-1+\gamma^2}},$$
where $a>1$, $0<\gamma\leq1$ is determined by an optimization problem over the null space. When  $\gamma< 1$ this bound is greater than $\sqrt{\frac{a-1}{a}}$ from $\ell_1$ minimization. In addition, we also establish the bound on $\delta_{k}$ and show that it can be larger than the sharp one $1/3$ via $\ell_1$ minimization and also greater than $0.4343$ via weighted $\ell_1$ minimization under some mild cases. Finally, we achieve a modified iterative reweighted $\ell_1$ minimization (MIRL1) algorithm based on our selection principle of weight, and the numerical experiments demonstrate that our algorithm behaves much better than $\ell_1$  minimization and iterative reweighted $\ell_1$ minimization (IRL1) algorithm.
\end{abstract}


\section{Introduction}

\IEEEPARstart{W}{ith} dramatic advances in technology in recent years, various research fields, ranging from applied mathematics, computer science to engineering, have involved to recover some original $n$-dimensional but sparse data (e.g., signals and images) from linear measurement with dimension far fewer than $n$. This essential idea in terms of signal was first formulated as compressed sensing (CS)  by Donoho \cite{D8}, Cand$\grave{\textmd{e}}$s, Romberg and Tao \cite{CRT9} and Cand$\grave{\textmd{e}}$s and Tao \cite{CT1}. Since then myriads of researchers have been lured to this area as a consequence of its extensive applications in signal processing, communications, astronomy, biology, medicine, seismology and so forth, and thus brought fruitful theoretical results, see, e.g., survey papers \cite{BDE12,R} and monographs \cite{EK13,FR,NB}.

To acquire a sparse presentation $x\in \mathbb{R}^n$ of an underdetermined system of the form $\Phi x = b$, where $b\in \mathbb{R}^m$ is the available measurement and $\Phi \in \mathbb{R}^{m\times n}$ is a known measurement matrix (with $m<n$ ), the underlying model is the following $\ell_0$ \emph{minimization}
\begin{eqnarray}\label{l0} &&\textup{min} ~\|x\|_0,~~\textup{s.t.}~\Phi x=b,\end{eqnarray}
 where  $\|x\|_0$ is $\ell_0$-norm of the vector $x\in\mathbb{R}^n$, i.e., the number of nonzero
entries in $x$. Model (\ref{l0}) is a combinatorial optimization problem with a prohibitive complexity if solved by enumeration, and thus does not
appear tractable.

One common alternative approach is to solve (\ref{l0}) via its convex \emph{$\ell_1$ minimization}
\begin{eqnarray}\label{l1} &&\textup{min} ~\|x\|_1,~~\textup{s.t.}~\Phi x=b.\end{eqnarray}
The use of $\ell_1$ relaxation has become so widespread that it could arguably be considered the ¡°modern least squares¡±, see, e.g., \cite{BDE12,CWX2,CWX4,CZ,CZ4,CZ5,GN,ML,XX,Z,ZKX}.

Inspired by the efficiency of $\ell_1$ minimization, it is natural to ask, for example, whether a different
(but perhaps again convex) alternative to $ \ell_0$ minimization might also find the correct solution, but with a lower measurement requirement than $\ell_1$ minimization.

Earlier numerical experiments indicated that the reweighted $\ell_1$ minimization does outperform unweighted $\ell_1$ minimization in many situations \cite{CWB,DDFG,FR,NB,Z,ZL}. Therefore, reweighted $\ell_1$ relaxation for model (\ref{l0}) in decade have drawn large numbers of researchers to pay their attention on sparse signal recover due to its numerical computational advantage.

Because of this, there have been many researchers concentrated on studying the theoretical aspects of the weighted $\ell_1$ minimization \cite{FMSY,KXAB}. In this paper, as a sequence, we also consider the theoretical properties of the \emph{weighted $\ell_1$ minimization}
\begin{eqnarray}\label{lw}&& \textup{min} ~\|\omega\circ x\|_1,~~\textup{s.t.}~\Phi x=b,\end{eqnarray}
where $\circ$ denotes the Hadamard product, that is $\|w\circ x\|_1=\sum\omega_i|x_i|$, and $0<\omega_i\leq1,~i=1,\cdots,n.$ Here if we let $\omega$ as
\begin{numcases}{\omega_i=}
1-\epsilon,~~ i\in T,\nonumber\\
1,~~~~~~i\in T^C, \nonumber\end{numcases}
where $0<\epsilon<1$, $T$ is the subset of $\left\{1,2,\cdots,n\right\}$ and $T^C$ notates the complementary set of $T$ in $\left\{1,2,\cdots,n\right\}$, then (\ref{lw}) can be written as
\begin{eqnarray}\label{lw1}&& \textup{min} ~\|x\|_1-\epsilon\|x_T\|_1,~~\textup{s.t.}~\Phi x=b,\end{eqnarray}
where $x_T\in\mathbb{R}^{n}$ denotes the vector equals to $x$ on an index set $T$ and zero elsewhere. It is evident that model (\ref{lw1}) is a specific form of the difference of two convex functions programming (DC programming, see, e.g., \cite{GRC}).

For the sake of convenience to illustrate, we can draw a picture (see, Fig \ref{fig1}, where $\Phi$ and $b$ are given as Example \ref{ex1}) to comprehend the advantage of weighted $\ell_1$ minimization what is absent in $\ell_1$ minimization.
\vspace{2mm}

\begin{figwindow}[0,c,%
 \fbox{\includegraphics[width=85mm]{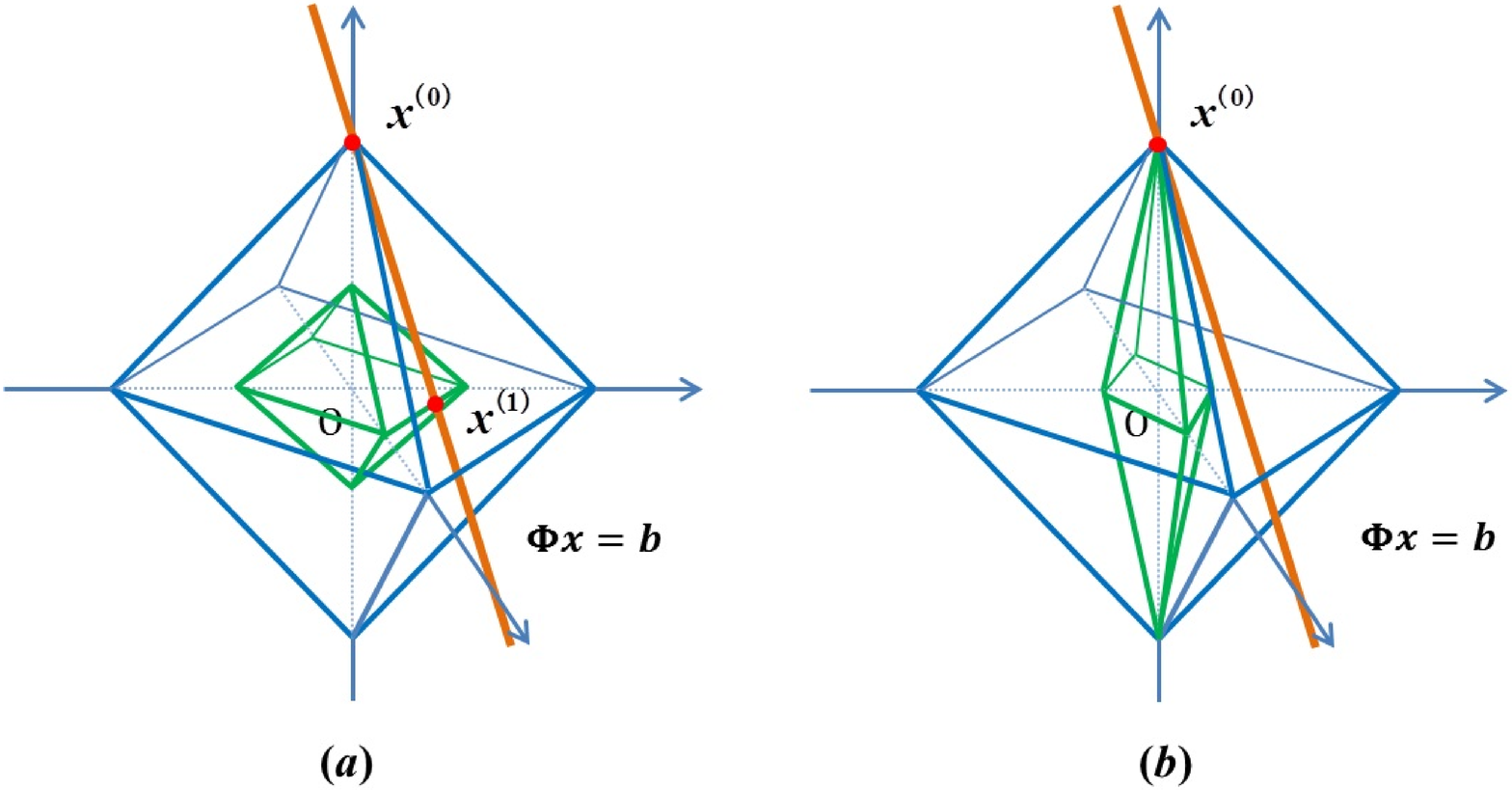}},%
                  { Some cases that $\ell_1$ minimization will fail to recover the sparse signal while exact recovery can be succeeded via weighted $\ell_1$ minimization. (a) Sparse signal $x^{(0)}=(0,0,2)^{T}$, feasible set $\Phi x = b$, and in $\ell_1$ ball there exists an $x^{(1)}=(\frac{3}{4},\frac{3}{4},0)^{T}$ but $\|x^{(1)}\|_0> \|x^{(0)}\|_0$. (b) In weighted $\ell_1$ ball, there does not exist an $x\neq x^{(0)}$ such that $\|x\|_0\leq\|x^{(0)}\|_0$.}\label{fig1}]
\end{figwindow}
\vspace{2mm}
Now let us recollect the theoretical properties of the standard $\ell_1$ minimization (\ref{l1}). We know that the \emph{null space property} (NSP) is the necessary and sufficient condition for (\ref{l1}) to reconstruct the system $b=\Phi x$ exactly \cite{DE,GN,ZY}. The NSP is recalled as follows.
\begin{Definition}[NSP]\label{de1}  A matrix $\Phi\in\mathbb{R}^{m\times n} $ satisfies the null space property of order $k$ if for all subsets $S\in\mathcal{C}_{n}^{k}$ it holds
\begin{eqnarray}\label{nsp}\left\|  h_{S}\right\|_1<\left\| h_{S^C}\right\|_1\end{eqnarray}
 for any $h\in \mathcal{N}(\Phi)\setminus\{0\}$, where $\mathcal{N}(\Phi)=\{h\in\mathbb{R}^{n}\large|~\Phi h=0\}$ and $\mathcal{C}_{n}^{k}=\left\{S\subset\{1,2,\cdots,n\}~\large|~|S|=k\right\}$ .
\end{Definition}

Another most popular sufficient condition for exact sparse recovery is related to the \emph{Restricted Isometry Property} (RIP) originated by Cand$\grave{\textmd{e}}$s and Tao~\cite{CT1}.
\begin{Definition}[RIP]\label{de2}  For $k \in\{ 1, 2,\cdots,n\}$, the restricted isometry constant is the smallest positive number $\delta_k$ such that
\begin{eqnarray}\label{derta} (1-\delta_k)\|x\|_2^2\leq\|\Phi x\|_2^2 \leq(1+\delta_k)\|x\|_2^2\end{eqnarray}
holds for all $k$-sparse vector $x\in\mathbb{R}^n$, i.e., $\|x\|_0\leq k$.
\end{Definition}
Current upper bounds on the restricted isometry constants (RICs) via  $\ell_1$  minimizations for exact signal recovery were emerged in many studies \cite{AS,CWX2,CZ,CZ4,CZ5,ML,ZKX}, such as $\delta_{2k}< 0.5746$ jointly with $\delta_{8k}< 1$ \cite{ZKX}, an improved bound $\delta_{2k}<\frac{4}{\sqrt{41}}$ \cite{AS}, sharp ones $\delta_{2k}<\frac{\sqrt{2}}{2}$ \cite{CZ5} and $\delta_{k}<\frac{1}{3}$ \cite{CZ}. As for the weighted $\ell_1$ minimization, literature \cite{FMSY} presented us the upper bound on $\delta_{k}$ might be $\delta_{k}<0.4343$ under some cases.

The main contributions of this paper are four aspects:
\begin{itemize}
 \item The WNSP, one necessary and sufficient condition for exact recovery via the weighted $\ell_1$ minimization, has been established, and then we comprehend its weakness compared to the standard NSP by illustrating some examples.
  \item  We then prove that the RIC bound by weighted $\ell_1$ minimization is
  $$\delta_{ak}< \sqrt{\frac{a-1}{a-1+\gamma^2}},$$ where $a>1$, $0<\gamma\leq1$ is determined by an optimization problem over the null space of $\Phi$. When $\gamma<1$ this bound is greater than $\sqrt{\frac{a-1}{a}}$ from $\ell_1$ minimization, which signifies that the scale of the undetermined measurement matrices, satisfying the RIP to ensure exact recovery via weighted $\ell_1$ minimization, is larger than those via $\ell_1$ minimization.
  \item The bound on $\delta_{k}$ has been given as well, and the result shows that it can be larger than the sharp one $\frac{1}{3}$ via $\ell_1$ minimization and also greater than $0.4343$ under some mild cases.
  \item Finally, based on the RIC theory, we achieve a modified iterative reweighted $\ell_1$ minimization (MIRL1) algorithm by establishing an effective way to add the weights. The numerical experiments demonstrate our method behaves much better than non-weighted $\ell_1$ minimization and iterative reweighted $\ell_1$ minimization (MIRL1) algorithm.
\end{itemize}

The organization of this paper is as follows. In Section II, we establish the necessary and sufficient condition for exact recovery via weighted $\ell_1$ minimization. And then by acquiring the upper bound on RIC, we set up another sufficient condition and give some examples to illustrate our results in Section III. The design of modified iterative reweighted $\ell_1$ minimization algorithm and numerical experiments will be presented in Section IV. We make a conclusion in Section V and give all of proofs in the last section.

\section{Weighted Null Space Property}
The \emph{Null Space Property} (NSP) is the necessary and sufficient condition for relaxation (\ref{l1}) to exactly recover problem (\ref{l0}). We know that $\mathcal{N}(\Phi)$ is a convex cone, also a subspace in $\mathbb{R}^{n}$, which means we can concentrate all information on one of its bases. Here we define a subset $\mathcal{N}_\varsigma$ from $\mathcal{N}(\Phi)$ by
\begin{eqnarray}\label{n1}\mathcal{N}_\varsigma=\{h\in\mathbb{R}^{n}\large|~h\in \mathcal{N}(\Phi), \|h\|_1=\varsigma\},\end{eqnarray}
where $\varsigma>0$ and any $\mathcal{N}_\varsigma$ is a base of $\mathcal{N}(\Phi)$. Since the fact of $\mathcal{N}(\Phi)\setminus\{0\}=\underset{\varsigma>0}{\bigcup}\mathcal{N}_\varsigma$, we can cast the NSP as follows.
\begin{Definition}\label{de3} A matrix $\Phi\in\mathbb{R}^{m\times n} $ satisfies the null space property of order $k$ if for all subsets $S\in\mathcal{C}_{n}^{k}$ it holds
\begin{eqnarray}\label{nsp}\left\|  h_{S}\right\|_1<\left\| h_{S^C}\right\|_1\end{eqnarray}
 for any $h\in \mathcal{N}_1$.
\end{Definition}
\begin{Lemma}\label{lemma00}Definition \emph{\ref{de1}} is equivalent to Definition \emph{\ref{de3}}.\end{Lemma}
Likewise, we give a \emph{Weighted Null Space Property} (WNSP) for the weighted $\ell_1$ minimization (\ref{lw}). Actually, literature \cite{KXAB} has already shown us the WNSP, here we will formulate it based on our Definition \ref{de3}.
\begin{Definition}[WNSP] For a given weight $\omega\in\mathbb{R}^{ n} $,  a matrix $\Phi\in\mathbb{R}^{m\times n} $ satisfies the weighted null space property of order $k$ if for all subsets $S\in\mathcal{C}_{n}^{k}$ it holds
\begin{eqnarray}\label{wnsp}\left\|\omega\circ h_{S}\right\|_1<\left\|\omega\circ h_{S^C}\right\|_1\end{eqnarray}
 for any $h\in \mathcal{N}_1.$
\end{Definition}
Similarly, by Lemma \ref{lemma00}, the WNSP that is built up on subset $\mathcal{N}_1$ also holds for the entire space $\mathcal{N}(\Phi)\setminus\{0\}$. For clearness, we will concentrate all sequent analysis on the subset $\mathcal{N}_1$ instead of $\mathcal{N}(\Phi)\setminus\{0\}$. Based on the WNSP we have the following recovery result linked to the weighted $\ell_1$ minimization.
\begin{Theorem}\label{Theorem1}
Every $k$-sparse vector $\hat{x}\in\mathbb{R}^n$ is the unique solution of the weighted minimization $(\ref{lw})$ with $b=\Phi\hat{x}$ if and only if $\Phi$  satisfies the WNSP of order $k$.\end{Theorem}

Now let us utilize two examples, which both satisfy the WNSP we defined while does not content the NSP, to illustrate the WNSP is a weaker exact recovery condition than the NSP.
\begin{Example}\label{ex1} Let the measurement matrix $\Phi$ and observation vector $b$ be given as
 $$\Phi=\left(
   \begin{array}{ccc}
     4/5 & 0 & 3/10 \\
     0 & 4/5 & 3/10 \\
   \end{array}
 \right),~~b=\left(
               \begin{array}{c}
                 3/5 \\
                 3/5 \\
               \end{array}
             \right)
 .$$\end{Example}
\noindent Clearly, the unique solution of $\ell_0$ and $\ell_1$ minimizations are $x^{(0)}=(0,0,2)^{T}$ and $x^{(1)}=(\frac{3}{4},\frac{3}{4},0)^{T}$ respectively. If setting $\omega_2=\omega_1, \omega_3<\frac{3}{4}\omega_1$ and $\omega_1\in(0,1]$, we can verify that $x^{(0)}$ is also the unique solution of the weighted $\ell_1$ minimization (For more clearness, one can see Fig \ref{fig1}).

For any $h=(h_1,h_2,h_3)^{T}\in \mathcal{N}_1$, by directly calculating, we have $h=(\frac{3}{8}h_3,\frac{3}{8}h_3,-h_3)^{T}$ with $h_3=4/7$ . Then for all subset $S\in\mathcal{C}_{3}^{1}$ and the given $\omega$ it holds
\begin{eqnarray}\left\|\omega\circ h_{S}\right\|_1<\left\|\omega\circ h_{S^C}\right\|_1.\nonumber\end{eqnarray}
From Theorem \ref{Theorem1}, the weighted $\ell_1$ minimization can exactly recover the sparsest solution of $\Phi x=b$. It is worth mentioning that this $\Phi$ does not satisfy the NSP due to $|h_3|\nless|\frac{3}{4}h_3|=|h_1|+|h_2|$ and thus the standard $\ell_1$ minimization will fail to exact recovery.\qed

 \begin{Example}\label{ex2} Let the measurement matrix $\Phi$ and observation vector $b$ be given as
 $$\Phi=\left(
   \begin{array}{ccccc}
     3/4 & -1/2 & 3/8&1/2&-1/4 \\
     3/4 & -1/2 & -1/8&1/2&0 \\
     0 & 1/4 & 3/8&-1/8&-3/8 \\
   \end{array}
 \right),$$
 and $b=\left(1/2 ,1/2,-1/8\right)^T$.\end{Example}
 \noindent It is easy to verify the unique solution of $\ell_0$ and $\ell_1$ minimizations are $x^{(0)}=(0,0,0,1,0)^{T}$ and $x^{(1)}=(\frac{1}{3},-\frac{1}{2},0,0,0)^{T}$ respectively. If setting $\omega_2=\frac{2}{3}\omega_1, \omega_4=\frac{1}{2}\omega_1,\omega_3=\omega_5=\omega_1$ and $ \omega_1\in(0,1]$, we can verify that $x^{(0)}$ is also the optimal solution of the weighted $\ell_1$ minimization.

For any $h=(h_1,h_2,h_3,h_4,h_5)^{T}\in \mathcal{N}_1$, by directly calculating, $h$ with $\|h\|_1=1$ has the following formation
  \begin{eqnarray}h=\left(\frac{-8 h_2+13 h_5}{12},h_2,\frac{h_5}{2},\frac{4h_2- 3 h_5}{2},h_5\right)^{T}.\nonumber\end{eqnarray}
 Then for all subset $S\in\mathcal{C}_{5}^{1}$ and the given $\omega$ it holds
 \begin{eqnarray}\left\|\omega\circ h_{S}\right\|_1<\left\|\omega\circ h_{S^C}\right\|_1.\nonumber\end{eqnarray}
From Theorem \ref{Theorem1}, the weighted $\ell_1$ minimization can exactly recover the sparsest solution of $\Phi x=b$. It is worth mentioning that this $h$ does not satisfy the NSP due to $|2h_2|\nless|\frac{2}{3}h_2|+|h_2|$ when $|h_5|=0$ and thus the standard $\ell_1$ minimization will fail to exact recovery.\qed

\section{Restricted Isometry Property}
In this section, we will study a sufficient condition, \emph{Restricted Isometry Property} (RIP), for the weighted $\ell_1$ minimization (\ref{lw}) to exactly recover model (\ref{l0}).  The first lemma about the sparse representation of a polytope established by  Cai and Zhang \cite{CZ5} will be very useful to prove our result, whose description is recalled bellow.

\begin{Lemma}\label{lemma0}
For a positive number $\alpha$ and a positive integer $s$, define the polytope $T(\alpha, s)\subset \mathbb{R}^n$ by
$$T(\alpha,s) = \left\{v\in\mathbb{R}^n \large ~|~ \|v\|_\infty \leq\alpha, \|v\|_1\leq s\alpha\right\}.$$
For any $v\in\mathbb{R}^n$, define the set $U(\alpha, s, v)\subset \mathbb{R}^n$ of sparse vectors by
 \vspace{-6mm}
\begin{flushright}\begin{eqnarray}U(\alpha,s, v) =\{u\in\mathbb{R}^n \large~|~{\wuhao\emph{supp}}(u)\subseteq {\wuhao\emph{supp}}(v),\|u\|_0\leq s,\nonumber\\
\|u\|_1 = \|v\|_1,\|u\|_\infty \leq \alpha\}.\nonumber\end{eqnarray}\end{flushright}
Then $v\in T(\alpha,s)$ if and only if $v$ is in the convex hull of $U(\alpha, s, v)$. In particular, any $v\in T(\alpha,s)$ can be expressed as $v=\sum_{i=1}^{N}\lambda_iu_i,$ where $N\geq1$ is an integer and~$$0\leq\lambda_i\leq1,\sum_{i=1}^{N}\lambda_i=1,u_i\in U(\alpha, s, v),i=1,2,\cdots,N.$$
\end{Lemma}
In order to analyze and acquire the upper bounds on RIC, we first design a way of weighing and introduce some notations. We will see that the way of weighing plays a crucial role in obtaining our main results in this section. Let $T_0$ and $\widehat{h}$ be the optimal solution of the following model
\begin{eqnarray}\label{t0} (T_0,\widehat{h}):=\underset{T\in\mathcal{C}_{n}^{k},h\in\mathcal{N}_1}{\textmd{argmax}}\| h_{T}\|_1.\end{eqnarray}
\noindent For a constant $0<\gamma\leq1$, we define $\omega$ based on $T_0$ as
\begin{numcases}{\omega_{i}=}
\label{wt0}\gamma,~~~ i\in T_0,\\
\label{wt0c}1,~~~i\in T_0^{C},\nonumber\end{numcases}
where $T_0^{C}$ is the complementary set of $T_0$ in $\left\{1,2,\cdots,n\right\}$.

From (\ref{t0}) and (\ref{wt0}) we manage to decide the locations where the entries should be added a weight $\gamma$, which implies that the way to define the weight $\omega$, in a sense, give us a hint to acquire a meaningful and practical weight to pursue the sparse solution, despite we can not easily value those weights since (\ref{t0}) is a combinational optimization problem.
\begin{Lemma}\label{lemma1}Let $T_0$ and $\widehat{h}$ be defined as $(\ref{t0})$. If $T_0$ uniquely exists, then there exists $\omega$ defined as $(\ref{wt0})$ with $0<\gamma<1$ such that
  \begin{eqnarray}\label{maxw}\|\omega\circ \widehat{h}_{T_0}\|_1=\underset{T\in\mathcal{C}_{n}^{k},h\in\mathcal{N}_1}{\max}\|\omega\circ h_{T}\|_1.\end{eqnarray}
  If $T_0$ exists but not uniquely, then $\omega$ defined as $(\ref{wt0})$ with $\gamma=1$ satisfies $(\ref{maxw})$.\end{Lemma}

\begin{Lemma}\label{lemma2}Let $T_0$ and $\widehat{h}$ be defined by $(\ref{t0})$. For the given $\omega$ as $(\ref{wt0})$, if
\begin{eqnarray}\label{t0nsp} \|\widehat{h}_{T_0^C}\|_1>\gamma\| \widehat{h}_{T_0}\|_1\end{eqnarray}
holds, then the WNSP of order $k$ is followed.
\end{Lemma}

Now we give our result associated with getting the upper bounds on RICs: $\delta_{ak}$ for some $a>1$ and $\delta_{k}$.
\begin{Theorem}\label{Theorem2}For the given $\gamma$ and $\omega$ as $(\ref{t0})$ and $(\ref{wt0})$, if
\begin{eqnarray}\label{delta1}\delta_{ak}< \sqrt{\frac{a-1}{a-1+\gamma^2}}\end{eqnarray} holds for some $a>1$, then  each $k$ sparse minimizer $\hat{x}$ of the weighted $\ell_1$ minimization $(\ref{lw})$ is the solution of $(\ref{l0})$.
\end{Theorem}

From (\ref{delta1}) we list TABLE \ref{tab1} by taking different values of $\gamma.$
\begin{table}[h]
 \caption{Bounds on $\delta_{2k},\delta_{3k}$ and $\delta_{4k}$ with different cases.\label{tab1}}
{\renewcommand\baselinestretch{1.3}\selectfont
\hspace{0.9cm}{\xiaowuhao\begin{tabular}{ c| c c c}
\hline
$\gamma$&~~$\delta_{2k}$~~&~~$\delta_{3k}$~~&~~$\delta_{4k}$~~\\
\hline
~~~~1~~~~&~~~$\sqrt{2}/2$~~~&$~~~\sqrt{6}/3$~~~&~~~$\sqrt{3}/2$~~~\\
~~~~3/4~~~~&~~~0.800~~&~~~0.883~~&~~~0.917~~\\
~~~~1/2~~~~&~~~0.894~~&~~~0.942~~&~~~0.960~~\\
~~~~1/4~~~~&~~~0.970~~&~~~0.984~~&~~~0.989~~\\
\hline
\end{tabular}}
\par}
\end{table}

\begin{Theorem}\label{Theorem3}For the given $\gamma$ and $\omega$ as $(\ref{t0})$ and $(\ref{wt0})$, if
\begin{numcases}{\delta_{k}<}
\label{even}\frac{1}{1+2\lceil \gamma k\rceil/k},~~~\text{for even number}~ k\geq2,\\
\label{odd}\frac{1}{1+\frac{2\lceil \gamma k\rceil}{\sqrt{k^{2}-1}}},~~~~~ \text{for odd number}~ k\geq3,\end{numcases}
holds, where $\lceil a\rceil$ denotes the smallest integer that is no less than $a$, then each $k$ sparse minimizer $\hat{x}$ of the weighted $\ell_1$ minimization $(\ref{lw})$ is the solution of $(\ref{l0})$.
\end{Theorem}

From (\ref{even}) and (\ref{odd}) above, we list different RIC bounds on $\delta_{k}$ in TABLE \ref{tab2} by setting various $\gamma$ and $k$. From the table one cannot difficultly find that under some mild situation, the upper bounds are greater than $0.4343$ in \cite{FMSY}.
\begin{table}[h]
 \caption{Bounds on $\delta_{k}$ with different cases.\label{tab2}}
{\renewcommand\baselinestretch{1.35}\selectfont
\hspace{1.2cm}{\xiaowuhao
\begin{tabular}{ c| c c }
\hline
~$\gamma$&~~ $k\geq2$~\textmd{is even}~~~ & ~~~$k\geq3$~\textmd{is odd}~~~\\
\hline
1& $1/3$ ~~& ~~ $0.3203$~~~~\\
3/4& $3/8~(k\geq4)$ ~~& ~~$0.3797~(k\geq5)$~~~~\\
1/2 & $1/2~(k\geq2)$ ~~& ~~ $\sqrt{6}-2~(k\geq5)$~~~~\\
1/4 & $2/3~(k\geq4)$ ~~& ~~ $3-\sqrt{6}~(k\geq5)$~~~~\\
1/6 & $3/4~(k\geq6)$ ~~& ~~ $0.7101~(k\geq5)$~~~~\\
\hline
\end{tabular}}
\par}
\end{table}

To end this section we present two examples to illustrate Theorem \ref{Theorem2}, which both result in $\ell_1$ minimization failing to recover the sparsest solution of $\ell_0$ problem  while successful recovery with the help of the weighted $\ell_1$ minimization .
\begin{Example}\label{ex11}We consider Example $\emph{\ref{ex1}}$ again. \end{Example}
\noindent The optimal solution of $\ell_0$ is $x^{(0)}=(0,0,2)^{T}$. The unique solution of $\ell_1$ minimization is $x^{(1)}=(3/4,3/4,0)^{T}$. From $h=(\frac{3}{8}h_3,\frac{3}{8}h_3,-h_3)^{T}\in \mathcal{N}_1$ with $h_{3}=\frac{4}{7}$, $|h_{3}|$ is the largest entry of $h$, i.e. $T_0=\{3\}$ uniquely exists. Therefore by setting $\frac{3}{8}<\omega_3=\gamma<0.418$ and $\omega_1= \omega_2=1$, we have $\gamma\|h_{\{3\}}\|_1<\|h_{\{1,2\}}\|_1$, which means that $x^{(0)}$ is the unique solution of weighted $\ell_1$ minimization from Lemma \ref{lemma2} and Theorem \ref{Theorem1}.

 On the other hand, we directly calculate that $\delta_2=0.9224$ with $n=3,k=2$ by the following formula (see \cite{ML,ZKX})
 \begin{eqnarray}\label{del} \delta_k=\max_{S\in\mathcal{C}_{n}^{k}}\|\Phi_S^T\Phi_S-I_k\|,\end{eqnarray}
where $\|\cdot\|$ denotes the spectral norm of a matrix. Since $T_0$ uniquely exists and $\gamma<0.418$, it yields $\delta_2<0.9226$ from (\ref{delta1}) by taking $a=2,k=1$. Hence the $\ell_0$ minimization can be exactly reconstructed by the weighted $\ell_1$ minimization from our Theorem \ref{Theorem2}.\qed

 \begin{Example}\label{ex22}We consider Example \emph{\ref{ex2}} again. \end{Example}
\noindent The optimal solution of $\ell_0$ is $x^{(0)}=(0,0,0,1,0)^{T}$. The unique solution of $\ell_1$ minimization is $x^{(1)}=(\frac{1}{3},-\frac{1}{2},0,0,0)^{T}$. Since for any $h\in \mathcal{N}_1$, $h$ with $\|h\|_1=1$ has the formation
  \begin{eqnarray}h=\left(-2 h_2/3+13h_5/12,h_2,h_5/2,2h_2- 3 h_5/2,h_5\right)^{T}.\nonumber\end{eqnarray}
Simply calculating $(T_0,\widehat{h})=\underset{T\in\mathcal{C}_{5}^{1},h\in\mathcal{N}_1}{\textmd{argmax}}\| h_{T}\|_1$, it follows that
$$T_0=\{4\},~\widehat{h}=\left(-2 h_2/3,h_2,0,2h_2,0\right)^{T},~h_2=6/11,$$
which manifests that $T_0$ uniquely exists. Therefore by setting $\omega_4=\gamma=0.3$ and $\omega_1= \omega_2=\omega_3= \omega_5=1$, we have $\gamma\|h_{\{4\}}\|_1<\|h_{\{1,2,3,5\}}\|_1$, which means that $x^{(0)}$ is the unique solution of weighted $\ell_1$ minimization from Lemma \ref{lemma2} and Theorem \ref{Theorem1}.

 On the other hand, we compute $\delta_2=0.9572$ by (\ref{del}) with $n=5,k=2$. Since $T_0$ uniquely exists and $\gamma=0.3$, it yields $\delta_2<0.9578$ from (\ref{delta1}) by taking $a=2,k=1$. And thus the $\ell_0$ minimization can be exactly recovered via the weighted $\ell_1$ minimization from Theorem \ref{Theorem2}.\qed
\vspace{5mm}

To end this section, we will illustrate the rationality of the extra assumption that \emph{\textbf{$T_0$ defined by $(\ref{t0})$ uniquely exists}}, and the relationships between WNSP and NSP, WNSP and RIP via constructing some instances.

\begin{Remark} Although $T_0$ defined by $(\ref{t0})$ always exists but not uniquely sometimes. However, from Examples \emph{\ref{ex11}} and \emph{\ref{ex22}}, we can see the assumption that $T_0$ uniquely exists is actually not a strong assumption at least to a certain extent. Therefore our assumption is meaningful to achieve the goal of pursuing the sparse solution exactly.\end{Remark}

\begin{Remark}\emph{i)} WNSP is evidently an extension of NSP, and thus it is a weaker condition than NSP for exact revoery;\\
\emph{ii)} For some measurement matrices $\Phi$, there might be lots of $\omega$ satisfying WNSP but exist relatively fewer numbers of $\omega$ contenting $(\ref{delta1})$, which manifests the condition $(\ref{delta1})$ is stronger than $(\ref{wnsp})$ from WNSP.

\end{Remark}
We draw a graphic to illustrate the relationship between WNSP, NSP and RIP based on the statements above.
\vspace{2mm}

\begin{figwindow}[0,c,%
 \fbox{\includegraphics[width=70mm]{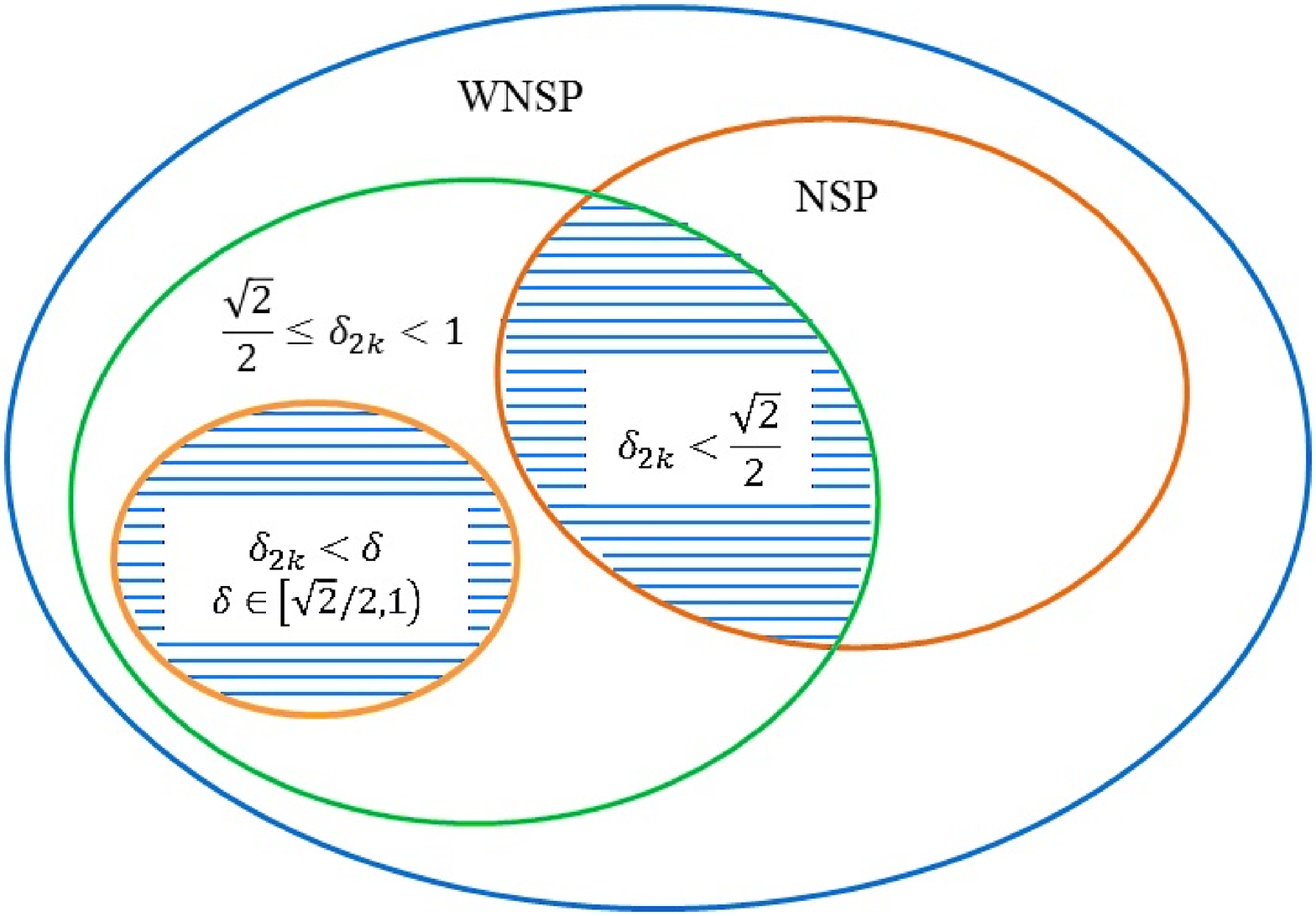}},%
                  {The relationship between WNSP, NSP and RIP, the dashed area denotes the scale of matrices that satisfy the RIP via weighted $\ell_1$ minimization. }\label{re}]
\end{figwindow}

\section{Numerical Experiments }

In this section, we will propose a modified iterative reweighted $\ell_1$ minimization (MIRL1) algorithm, where the weights are designed based on the theoretical results on the null space of $\Phi$. Simulation tests and signal experiments are also provided.
\subsection{Modified Iterative Reweighted $\ell_1$ Minimization}
Considering the following formula:
\begin{eqnarray}\label{lwa}\textup{min} ~\frac{1}{2}\|\Phi x-b\|_{2}^{2}+\mu\|\omega\circ x\|_1:=f(x),\end{eqnarray}
where $\mu>0$ is a penalty parameter. Let $L\geq\lambda_{\max}(\Phi^T\Phi)$. Then for any $x, y\in \mathbb{R}^{n}$, we have
\begin{eqnarray}&&\frac{1}{2}\|\Phi x-b\|_{2}^{2}+\mu\|\omega\circ x\|_1\nonumber\\
&\leq&\frac{1}{2}\|\Phi y-b\|_{2}^{2}+\langle\Phi^{T}(\Phi y-b),x-y\rangle+\frac{L}{2}\|x-y\|_{2}^{2}\nonumber\\
&&+\mu\|\omega\circ x\|_1\nonumber\\
&:=&F(x,y)\nonumber\end{eqnarray}
Evidently, for any $x, y \in\mathbb{R}^{n}$, we have
$$F(x, y) \geq f(x)~~\text{and }~~F(x, x) = f(x),$$
which means that $F$ is a majorization of $f$. Using this majorization function, we start with an initial iteration $x^{0}$ and update $x^{t}$ by solving
\begin{eqnarray}\label{so}x^{t+1}=\textup{argmin}_{x\in \mathbb{R}^{n}} ~F(x,x^{t}),\end{eqnarray}
which is equivalent to
\begin{eqnarray}\label{so1}x^{t+1}&=&\textup{argmin}_{x\in \mathbb{R}^{n}}~\frac{L}{2}\|x-\widetilde{x}^{t}\|_{2}^{2}+\mu\|\omega\circ x\|_1\nonumber\\
&=&\textmd{sign}(\widetilde{x}^{t})\circ\max\left\{|\widetilde{x}^{t}|-\frac{\mu}{L}\omega,0\right\}\end{eqnarray}
where
$$\widetilde{x}^{t}:=x^{t}-\frac{1}{L}\Phi^{T}(\Phi x^{t}-b),$$
$|x|=(|x_1|,|x_2|,\cdots,|x_n|)^T$ and $\textmd{sign}(x)$ denotes the signum function of $x$ . Here we need to indicate how to define the weight $\omega$. As we mentioned in Section III, since the weight $\omega$ is depended on the null space of $\Phi$, we take $T^{\tau}$ and $\omega^{\tau}$ as
\begin{eqnarray}\label{Tk}T^{\tau}&=&\textup{argmax}_{T\in \mathcal{C}_{n}^{k^{\tau}}}~\|(h^{\tau})_{T}\|_1,~~\tau=1,2,\cdots\end{eqnarray}
\begin{numcases}{\omega_{i}^{\tau}=}
\label{wk1}\left[\frac{|h^{\tau}_{i}|+\epsilon}{\max_{j\in (T^{\tau})^{C}}|h^{\tau}_{j}|}\right]^{q-1}~~~~~,i\in T^{\tau},~~\\
\label{wk2}~~~~~~~~~~1~~~~~~~~~~~~~~~~~~~,i\in (T^{\tau})^{C},\end{numcases}
where $$h^{\tau}=x^{\tau}-x^{\tau-1},~~~~k^{\tau}=|\textmd{supp}(x^{\tau})|$$ and $0<q\leq1, \epsilon>0$ is sufficiently small.
\begin{Remark}We simply interpret the weights as $(\ref{Tk})$--$(\ref{wk2})$. Simply verifying from $(\ref{Tk})$--$(\ref{wk2})$, we have $|h^{\tau}_{i}|\geq|h^{\tau}_{j}|,\forall~i\in T^{\tau},\forall~j\in (T^{\tau})^{C}$, and thus $|h^{\tau}_{i}|+\epsilon>\max_{j\in (T^{\tau})^{C}}|h^{\tau}_{j}|,\forall~i\in T^{\tau},$ which indicates $$0\leq\left[\frac{|h^{\tau}_{i}|+\epsilon}{\max_{j\in (T^{\tau})^{C}}|h^{\tau}_{j}|}\right]^{q-1}<1, ~~~\forall~i\in T^{\tau}.$$
\end{Remark}
\begin{Remark}To the best of our knowledge, the weights given by $(\ref{Tk})$--$(\ref{wk2})$ are different from those in the existing literature, see, e.g.,  $\cite{CWB,DDFG,FL,NB,ZL}$.  By partitioning the index set into parts $T^\tau$ and $(T^\tau)^C$
based on $h^\tau$, we endow the entries in two parts with corresponding weights. Moreover, we give weights $\omega^\tau$ from $h^\tau$
and no longer directly utilize $x^\tau$ to value the weight like $\omega_i^{\tau+1}=\frac{1}{|x_i^\tau|+\epsilon}$ in $\cite{CWB}$ or $\omega_i^{\tau+1}=\frac{1}{(|x_i^\tau|+\epsilon)^{1-q}}$, $q\in(0,1)$ in $\cite{FL}$, which can be uniformly written as
\begin{eqnarray}\label{wl1}\omega_i^{\tau+1}=\left[\frac{1}{|x_i^\tau|+\epsilon}\right]^{1-q},~~~~q\in[0,1).\end{eqnarray} \end{Remark}

\noindent Recall the well-known iterative reweighted $\ell_1$ minimization algorithm (IRL1) $\cite{CWB}$, we present the algorithm framework of our proposed modified version in TABLE \ref{tab22}.

\begin{table}[h]
 \caption{The framework of MIRL1 .\label{tab22}}
\renewcommand{\arraystretch}{1.5}\addtolength{\tabcolsep}{6pt}
\centering
\begin{tabular}{>{\small}l}
\hline
\textbf{Modified Iterative Reweighted $\ell_1$ Minimization (MIRL1)}        \\\hline
Initialize $x^{0}, \omega^{1}, M, \mu^{1}$ and $L\geq\sigma_{\max}(\Phi^T\Phi)$;        \\

\textbf{For} $\tau$=1: M\\
~~~~Initialize $x^{\tau,1}=x^{\tau-1}$;\\
~~~~\textbf{While} $\|x^{\tau,t+1}-x^{\tau,t}\|_2\geq\eta^{\tau}\max\{1,\|x^{\tau,t}\|_2\}$\\
~~~~~~~~ $\widetilde{x}^{\tau,t}=x^{\tau,t}-\frac{1}{L}\Phi^{T}(\Phi x^{\tau,t}-b)$; \\
~~~~~~~~ $x^{\tau,t+1}=\textmd{sign}\left(\widetilde{x}^{\tau,t}\right)\circ\max\left\{|\widetilde{x}^{\tau,t}|-\frac{\mu^{\tau}}{L}\omega^{\tau},0\right\}$. \\
~~~~\textbf{End}\\
~~~~Update $x^{\tau}=x^{\tau,t+1}$;\\
~~~~Update $\omega^{\tau+1}$ from $x^{\tau-1}, x^{\tau}$  based on (\ref{Tk}), (\ref{wk1}) and  (\ref{wk2}); \\
\textbf{End}\\
\hline
\end{tabular}\end{table}

Evidently, the framework of MIRL1 will go back to that of iterative reweighted $\ell_1$ minimization (IRL1) algorithm or  iterative $\ell_1$ minimization (IL1) algorithm if we update $\omega^{\tau+1}$ based on (\ref{wl1}) or $\omega^{\tau+1}=(1,1,\cdots,1)^{T}$, respectively.

\subsection{Computational Results: Exact Recovery}
We first consider the recovery without noise (exact recovery): $$y=\Phi x.$$
Before proceeding to the computational results, we need to define some notations and data sets. For convenience and clear understanding in the graph presentations and some comments, we use the notations: $L_1$, $WL_1$, $ML_1$ to represent the IL1 (namely derived from updating $\omega^{\tau+1}=(1,1,\cdots,1)^{T}$ in the framework), the IRL1 and MIRL1 respectively. Since the weight $\omega$ in (\ref{wk1})--(\ref{wk2}) and (\ref{wl1}) is associated with the parameter $q\in(0,1)$, we shortly write the methods as $WL_1(q=\varsigma)$ and $ML_1(q=\varsigma)$, particularly taking $\varsigma=0.1, 0.25, 0.5, 0.6, 0.75, 0.9$ in our whole numerical experiments. For each data set, the random matrix $\Phi$ and vector $b$ are
generated by the following matlab codes:
\begin{eqnarray}&&x_{\textmd{orig}} = \textmd{zeros}(n, 1),~~~~~ y = \textmd{randperm}(n),\nonumber\\
&&x_{\textmd{orig}}(y(1 : k)) = \textmd{randn}(k, 1),\nonumber\\
&&\Phi = \textmd{randn}(m, n),~~~~~~b = \Phi*x_{\textmd{orig}}. \nonumber\end{eqnarray}
The stopping criterias for the inner loops in our method are given by $\eta^{\tau}=\mu^{\tau}10^{-4},$ where parameters $\mu^{\tau}(\tau=1,2\cdots,8)$ are taken by
$$\mu^{\tau}\in\left\{1,1/5,(1/5)^2,\cdots,(1/5)^6, (1/5)^7\right\}\cdot\left\|\Phi^T b\right\|_{\infty},$$
which implies $\mu^1=(1/5)^0=1$ and $M=8$. We always initialize the start points $$x^0= \textrm{ones}(n,1),  ~~~~~~~\omega^1 = \textrm{ones}(n,1),~~~~~~~\epsilon=10^{-4}.$$

 \makeatletter
          \def\@captype{figure}
          \makeatother
{\flushleft\includegraphics[width=9.7cm]{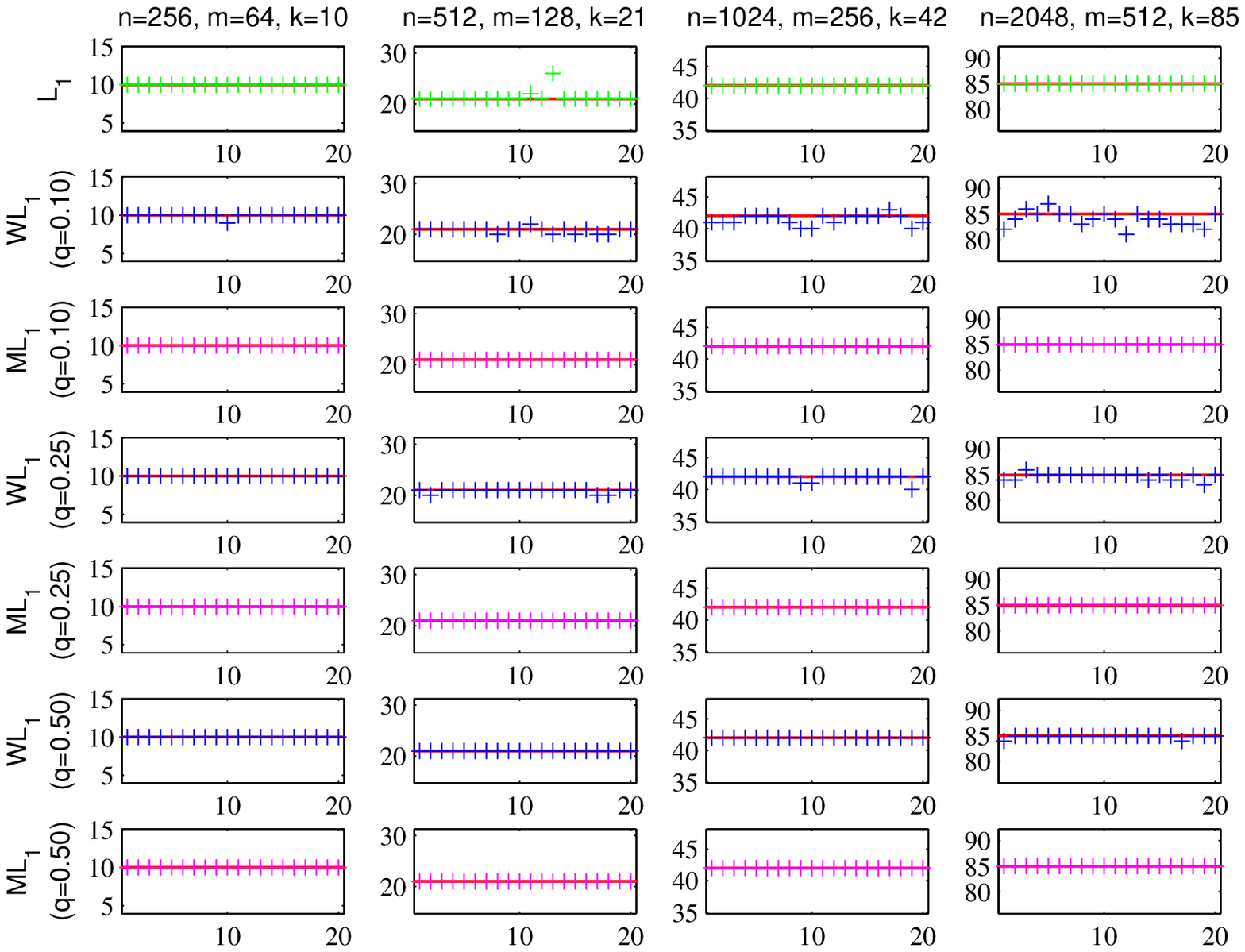}}
          \vspace{-9mm}
          \caption{Sparsity yielded by IL1, IRL1 and MIRL1 when $q=0.1, 0.25, 0.5$.\label{figs1}}
  \vspace{-4mm}
   \makeatletter
         \def\@captype{figure}
          \makeatother
          \begin{flushleft}
          \includegraphics[width=9.7cm]{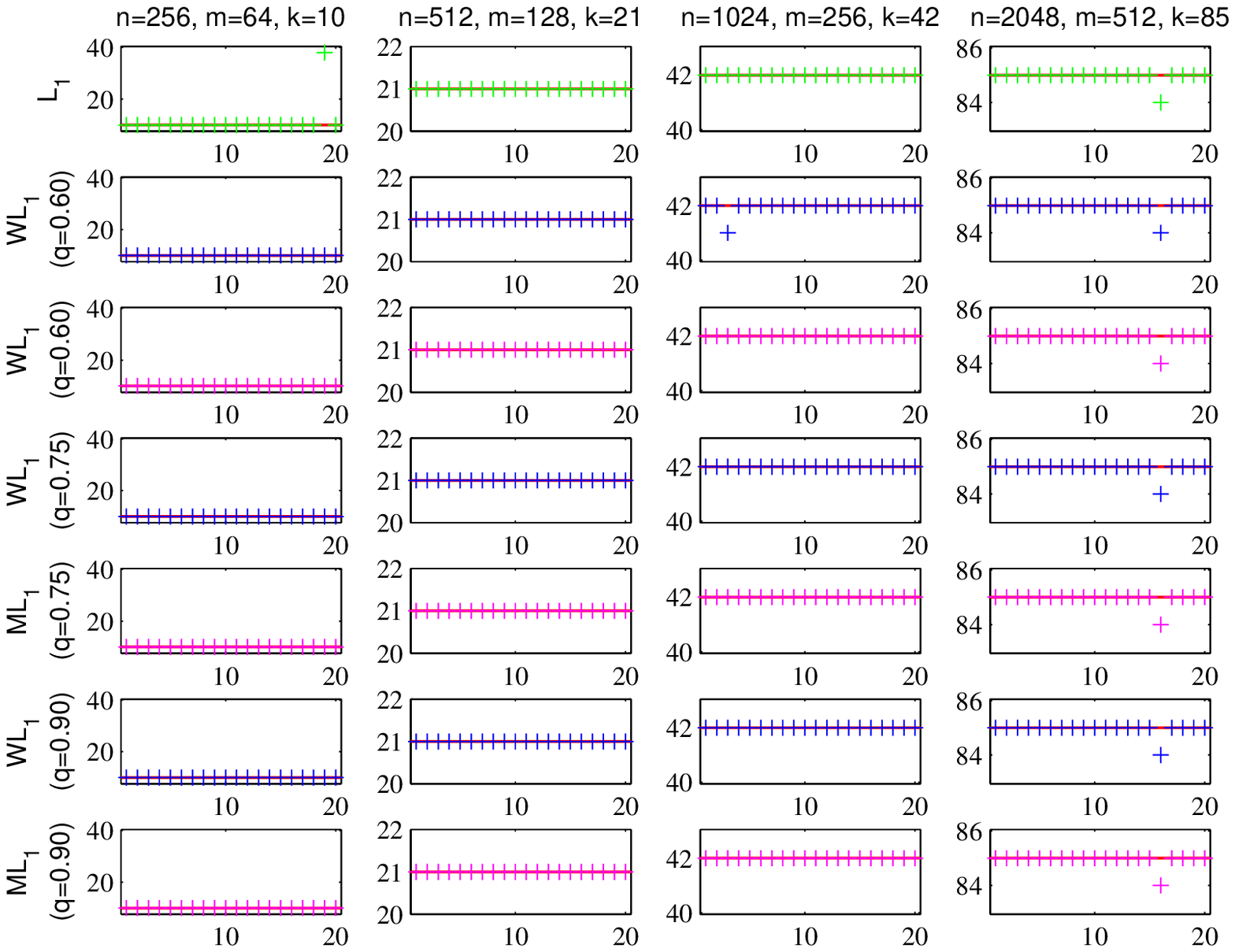}
          \end{flushleft}
         \vspace{-9mm}
          \caption{Sparsity yielded by  IL1, IRL1 and MIRL1 when $q=0.6, 0.75, 0.9$.\label{figs2}}
For each fixed $q=0.1, 0.25, 0.5 0.6, 0.75, 0.9$, we randomly generate $20$ samples and respectively apply IL1, IRL1 and MIRL1 algorithms to problem (\ref{lwa}). From Figs \ref{figs1} and \ref{figs2}, the red lines and the blue '$+$'s stand for the sparsity of $x_{\textmd{orig}}$ and recovered solutions, respectively. One can not be difficult to see the comments below.
 \begin{itemize}
   \item For any $q$, sparsity of the optimal solutions derived from MRIL1 is closer (almost equal) to the true sparsity than that from RIL1 and IL1.
   \item When $q=0.1, 0.25$, RIL1 performs relatively bad (also see TABLE \ref{tabsp}) while IL1 and MRIL1 still works steadily. Then with the increasing of $q(=0.5, 0.6, 0.75, 0.9)$, although there occasionally appears some bad cases, under such circumstance RIL1 and MRIL1 perform moderately better than IL1.
   \item Since there is no restriction on the CPU time to run the algorithms, IL1 has cost much more time than RIL1 and MRIL1, which contributes to obtaining the solutions whose sparsity is equal to the true one. Hence, that is reasonable for some $q$ the recovery effect is better for IL1 than that of RIL1.
 \end{itemize}

 \begin{center}
 \begin{table}[h]
 \caption{Sparsity yielded by  IL1, IRL1 and MIRL1 when $q=0.1, 0.5, 0.9$. \label{tabsp}}
{\renewcommand\baselinestretch{1.1}\selectfont
\begin{tabular}{ c| c| c c| c c| c c}\hline
\multicolumn{1}{c|}{\multirow{3}{*}{Sample}}&\multicolumn{7}{c}{$m=512,n=2048$, True Sparsity$=85$}\\\cline{2-8}
\multicolumn{1}{c|}{}&\multicolumn{1}{c|}{\multirow{2}{*}{$L_1$}}&\multicolumn{2}{c|}{$q=0.1$}&\multicolumn{2}{c|}{$q=0.5$}&\multicolumn{2}{c}{$q=0.9$}\\\cline{3-8}
\multicolumn{1}{c|}{}&\multicolumn{1}{c|}{}&\multicolumn{1}{c}{$WL_1$}&\multicolumn{1}{c|}{$ML_1$}&\multicolumn{1}{c}{$WL_1$}&\multicolumn{1}{c|}{$ML_1$}&
\multicolumn{1}{c}{$WL_1$}&\multicolumn{1}{c}{$ML_1$}\\
\hline
1&85&85&85&85&85&85&85\\
2&85&85&85&85&85&85&85\\
3&85&\textcolor[rgb]{0.98,0.00,0.00}{84}&85&85&85&85&85\\
4&85&85&85&85&85&85&85\\
5&85&\textcolor[rgb]{0.98,0.00,0.00}{83}&85&85&85&85&85\\
6&85&\textcolor[rgb]{0.98,0.00,0.00}{83}&85&85&85&85&85\\
7&85&85&85&\textcolor[rgb]{0.98,0.00,0.00}{84}&85&85&85\\
8&85&85&85&85&85&85&85\\
9&85&\textcolor[rgb]{0.98,0.00,0.00}{84}&85&85&85&85&85\\
10&85&\textcolor[rgb]{0.98,0.00,0.00}{84}&85&\textcolor[rgb]{0.98,0.00,0.00}{84}&85&85&85\\
\hline
\multicolumn{1}{c|}{}&\multicolumn{7}{c}{Average error $\|\Phi x-b\|_2$}\\\hline
($10^{-4}$)&12&7972&7&54&7&11&7\\\hline
\multicolumn{1}{c|}{}&\multicolumn{7}{c}{Average error $\|x-x_{\textmd{org}}\|_2$}\\\hline
($10^{-4}$)&0.35&213&	0.1&2&	0.11&	0.15&0.11\\\hline
\multicolumn{1}{c|}{}&\multicolumn{7}{c}{Average CPU time}\\\hline
(second)&5.67&3.00&3.69&3.14&3.38&3.17&3.16\\\hline
\end{tabular}\par}
\end{table}
\end{center}
\vspace{-8mm}
 \makeatletter
          \def\@captype{figure}
          \makeatother
          \begin{center}
          \includegraphics[width=9.1cm]{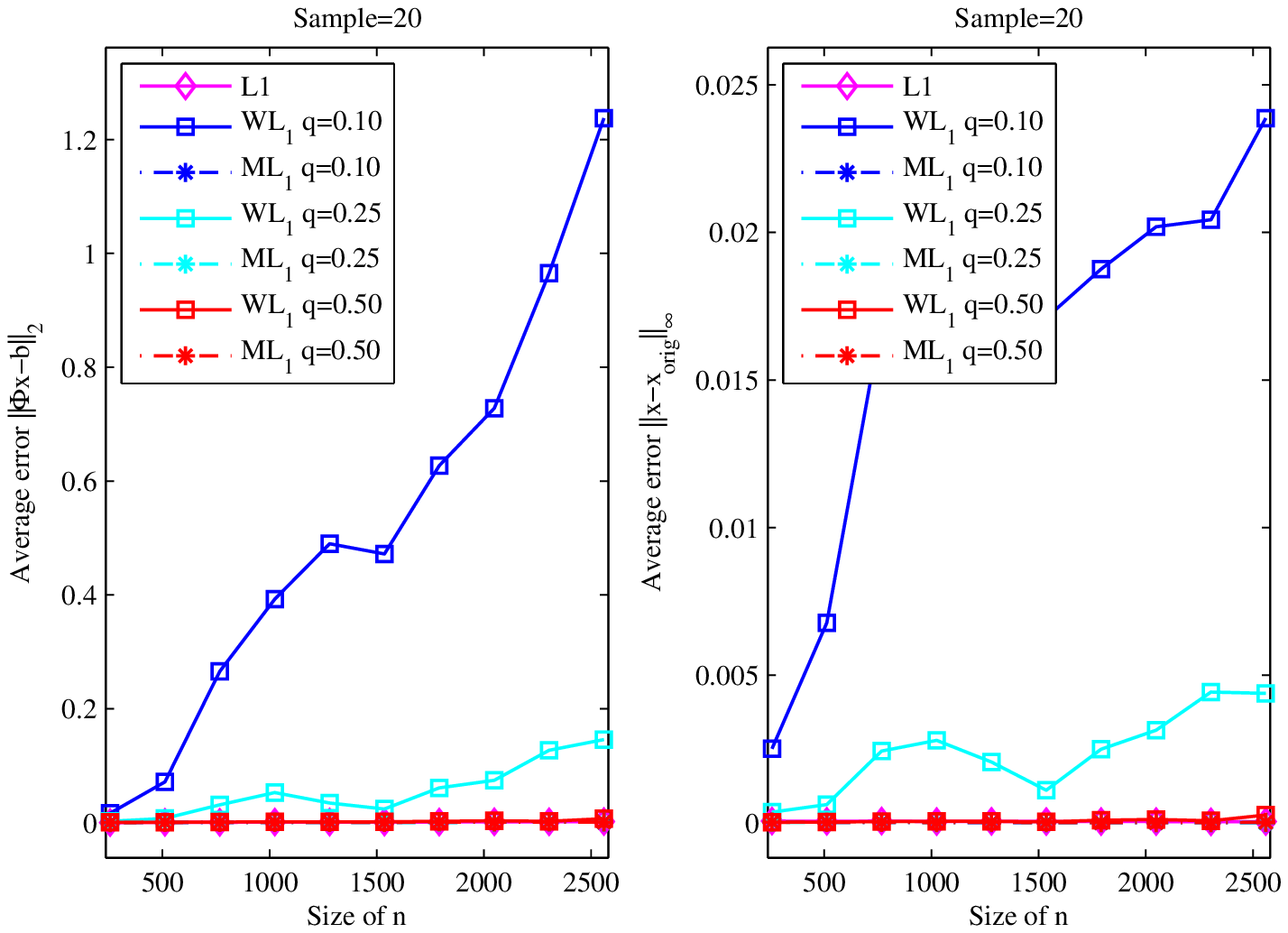}
          \vspace{-8mm}
          \caption{Error yielded by IL1, IRL1 and MIRL1 when $q=0.1, 0.25, 0.5$ .\label{fige1}}
\vspace{-2mm}
        \end{center}
         \makeatletter
          \def\@captype{figure}
          \makeatother
          \begin{center}
          \includegraphics[width=9.1cm]{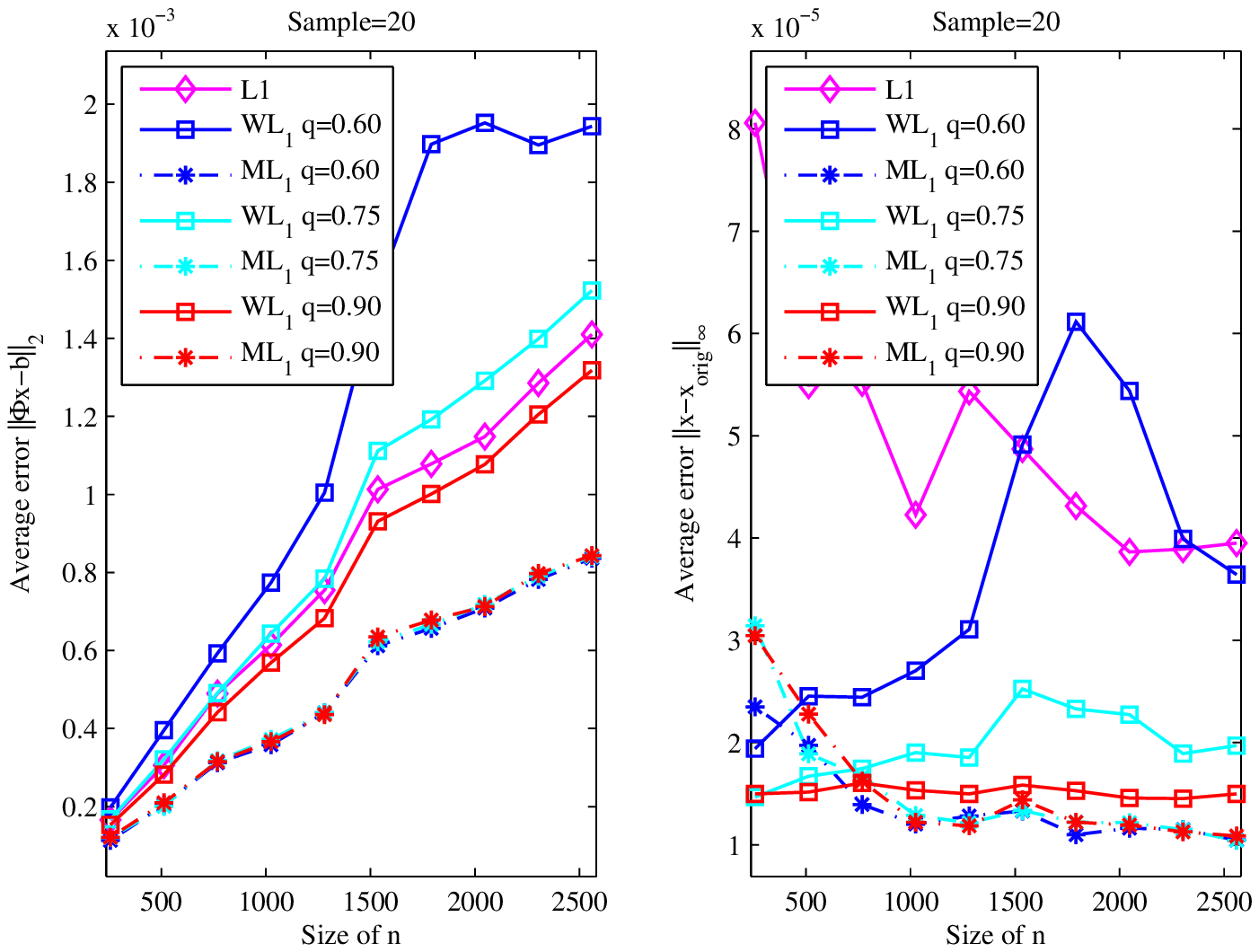}
          \vspace{-8mm}
          \caption{Error yielded by IL1, IRL1 and MIRL1 when $q=0.6, 0.75, 0.9$.\label{fige2}}
        \end{center}
From Figs \ref{fige1}--\ref{figt2}, several comments can be derived.
 \begin{itemize}
   \item For any $q$, the average errors $\|\Phi x-b\|_2$ (or $\|x-x_{\textmd{orig}}\|_\infty$) by MRIL1 are basically smaller than those of IL1 and  RIL1; Particularly, when $q=0.1$ or $0.25$, the average errors are much higher than others from MRIL1 and even from IL1. However with $q$ being no less than $0.5$, the average errors $\|x-x_{\textmd{orig}}\|_\infty$ almost become lower than IL1 but still higher than MRIL1;
   \item For RIL1, different $q$ would lead to some fluctuations of the average errors $\|\Phi x-b\|_2$ (or $\|x-x_{\textmd{orig}}\|_\infty$) which likely become intense with $n$ increasing, whilst MRIL1 would generate relatively stable errors' fluctuations regardless of $q$;
   \item For any $q$, Figs \ref{fige1} and \ref{fige2} see upward trends of the average errors from RIL1 with the ascend of $n$, whereas for any $q$ one can find that there are the downward trends of the average errors $\|x-x_{\textmd{orig}}\|_\infty$ resulted from MRIL1 when $n$ increases.
    \item The average CPU time generated by MRIL1 and RIL1 are basically equal, which are all much shorter than those from IL1. More specifically, all of them increase with the rise of dimension $n$, and the time spent by MRIL1 is slightly greater than that of RIL1 probably due to the computation of the weight (\ref{wk1})--(\ref{wk2}).
 \end{itemize}
\vspace{-4mm}
         \makeatletter
          \def\@captype{figure}
          \makeatother
          \begin{center}
          \includegraphics[width=9.cm]{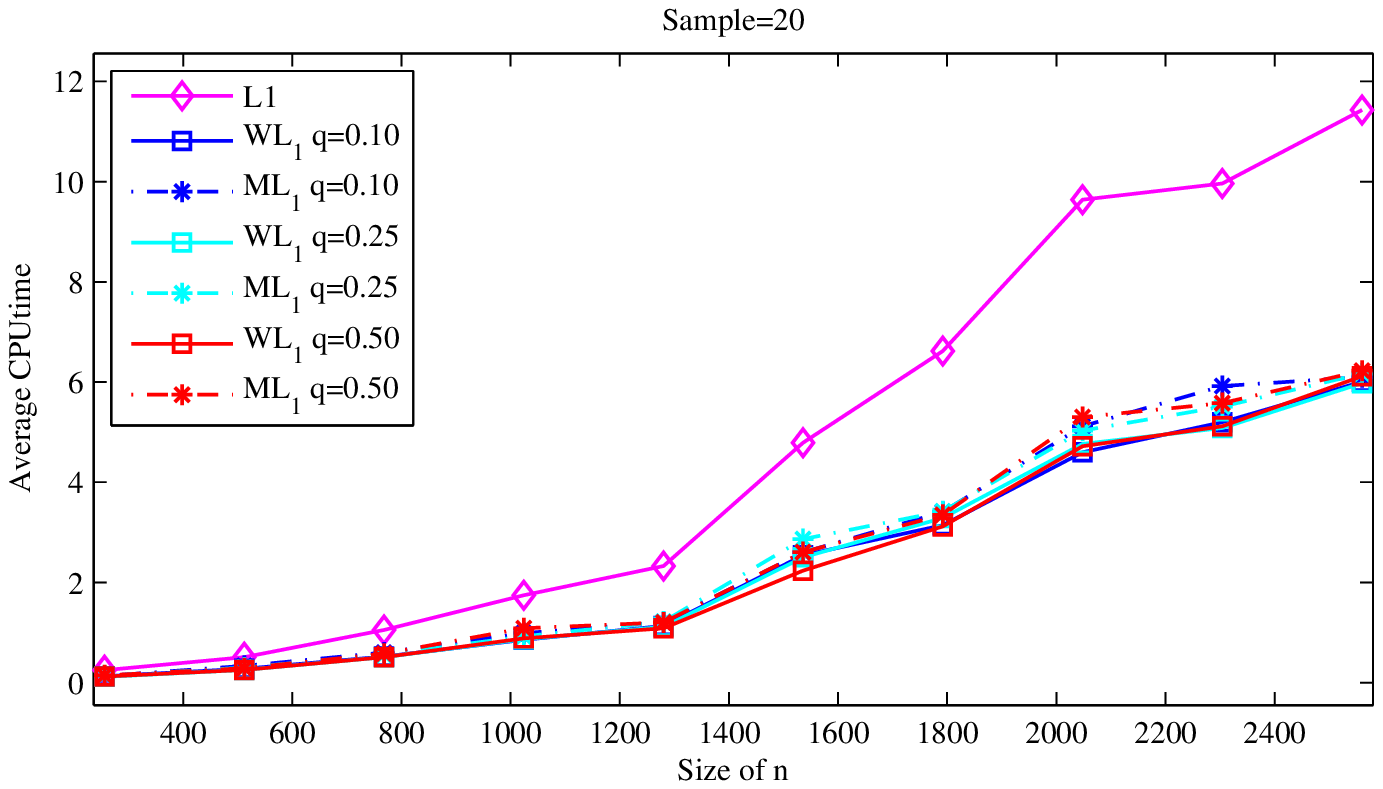}
          \vspace{-6mm}
          \caption{Time yielded by IL1, IRL1 and MIRL1 when $q=0.1, 0.25, 0.5$.\label{figt1}}
        \end{center}

\vspace{-6mm}
         \makeatletter
          \def\@captype{figure}
          \makeatother
          \begin{center}
          \includegraphics[width=9.cm]{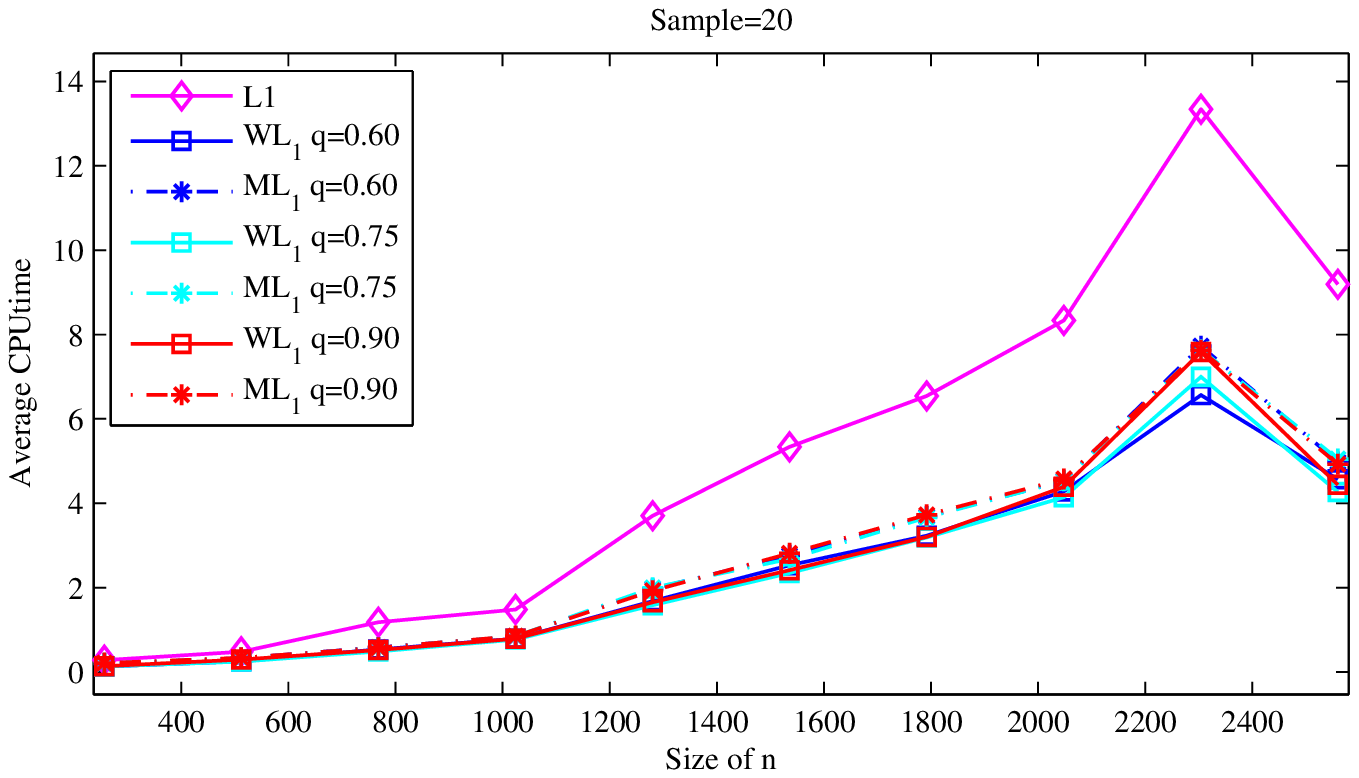}
          \vspace{-6mm}
          \caption{Time yielded by IL1, IRL1 and MIRL1 when $q=0.6, 0.75, 0.9$.\label{figt2}}
        \end{center}

From Figs \ref{figee1}--\ref{figee2} and TABLE \ref{tabav}, one can conclude the following comments.
\begin{itemize}
  \item For any $q\in\{0.1, 0.25, 0.5 0.6, 0.75, 0.9\}$, the average errors $\|\Phi x-b\|_2$ (or $\|x-x_{\textmd{orig}}\|_\infty$) by MRIL1 are quite small (almost reach from $10^{-3}$ to $10^{-5}$), which are much lower than those from IRL1 (most of whose values are greater than $10^{-3}$).
  \item Errors $\|\Phi x-b\|_2$ and $\|x-x_{\textmd{orig}}\|_\infty$ are basically equal for each $q$ when $n$ is fixed; In addition, the former increase while the latter decrease with the dimension $n$ rising;
  \item From TABLE \ref{tabav}, it is not of difficulty to see that our approach runs very fast, particularly when the sparsity $k=0.01n$, only 34.60 second is needed to pursue the sparse solution.
\end{itemize}
\vspace{-3mm}
   \makeatletter
          \def\@captype{figure}
          \makeatother
          \begin{center}
          \includegraphics[width=8.8cm]{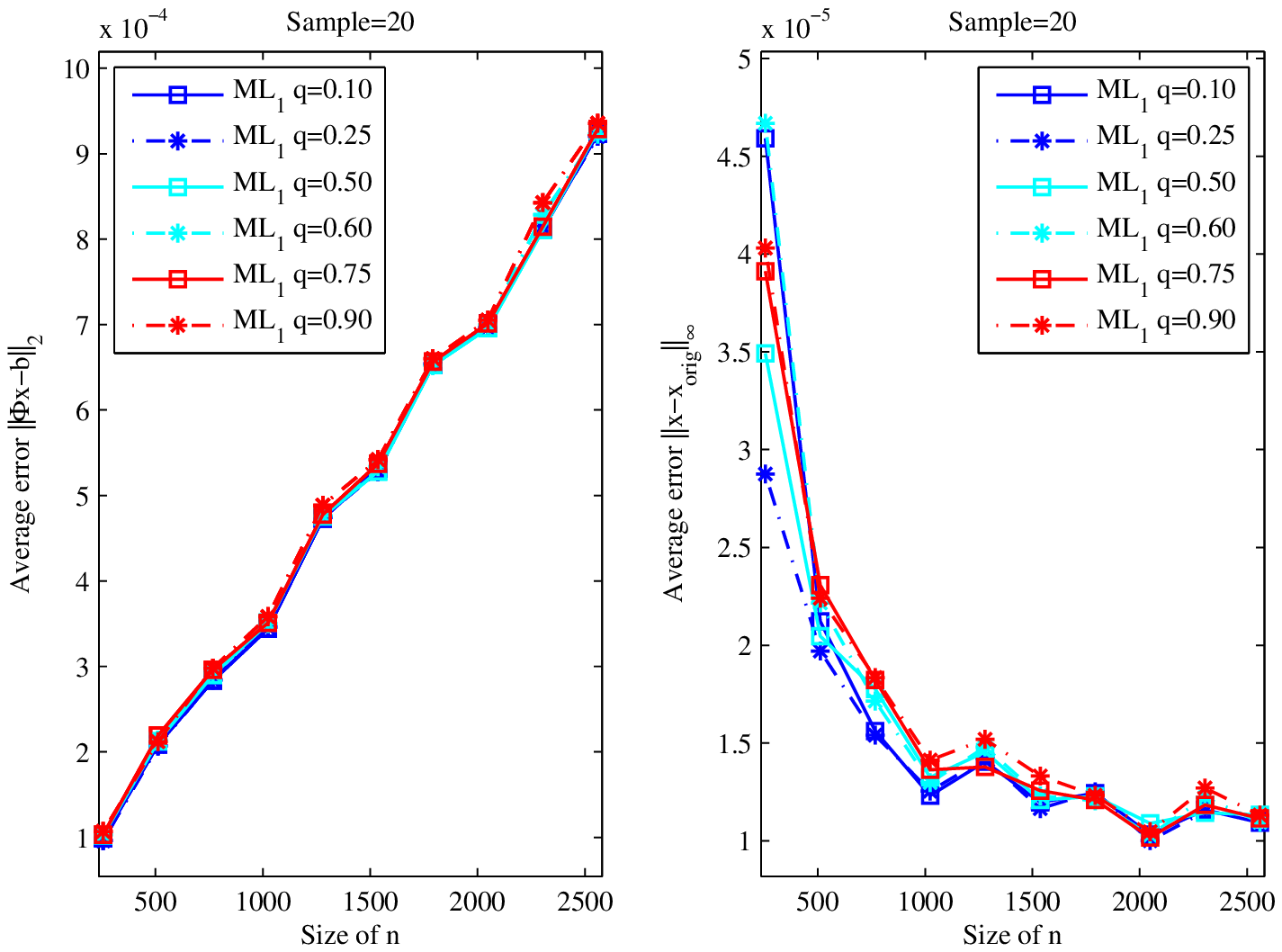}
          \vspace{-5mm}
          \caption{Error $\|\Phi x-b\|_2$ and $\|x-x_{\textmd{orig}}\|_\infty$ yielded by MIRL1.\label{figee1}}
        \end{center}
\vspace{-3mm}   \makeatletter
          \def\@captype{figure}
          \makeatother
          \begin{center}
          \includegraphics[width=8.9cm]{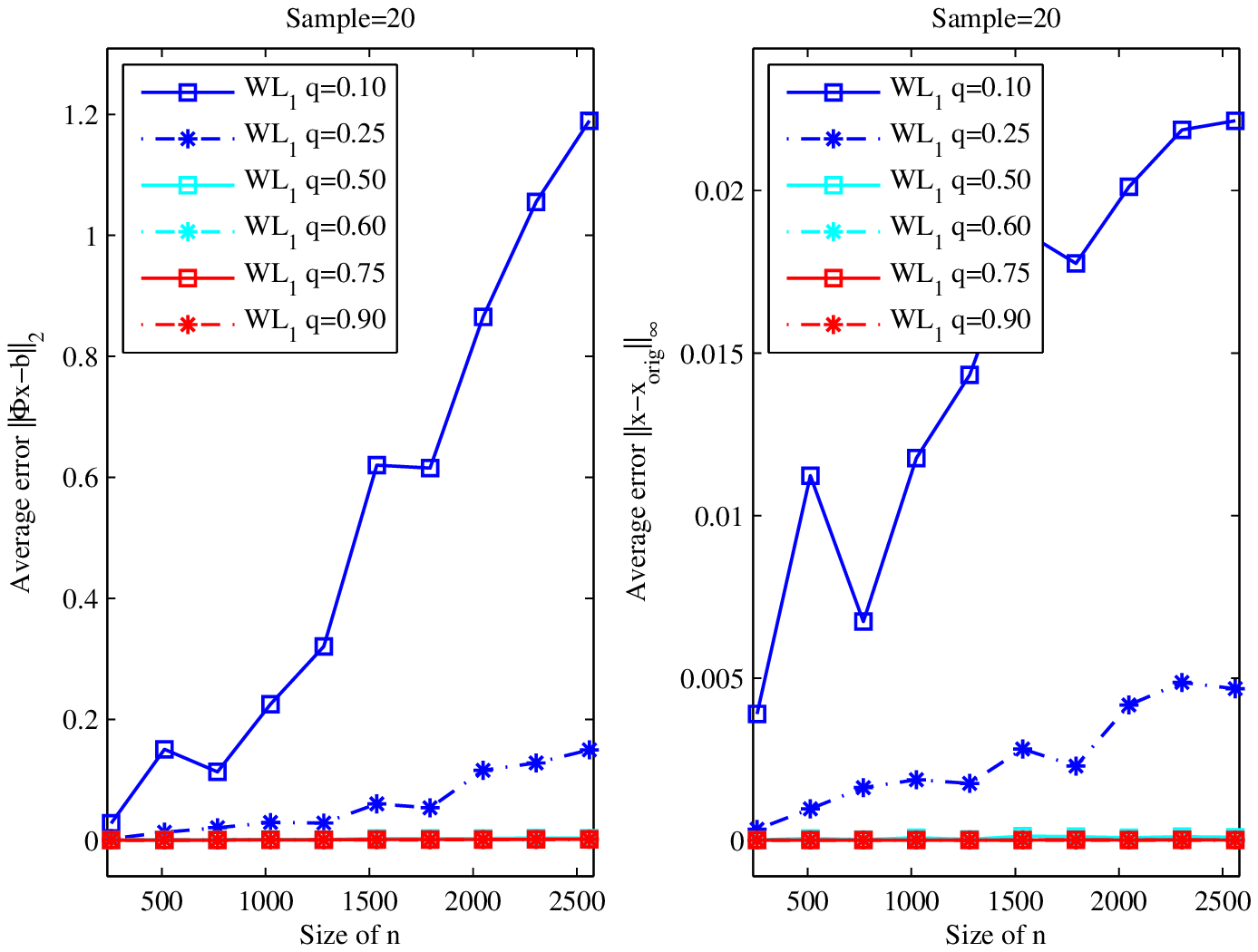}
          \vspace{-5mm}
          \caption{Error $\|\Phi x-b\|_2$ and $\|x-x_{\textmd{orig}}\|_\infty$ yielded by IRL1.\label{figee2}}
        \end{center}
\begin{center}
 \begin{table}[h]
 \caption{Average error and CPU time yielded by MIRL1 without noise . \label{tabav}}
{\centering{\renewcommand\baselinestretch{1.5}\selectfont
\begin{tabular}{ c| c| c c c }\hline
&$n$&~~~$\|\Phi x-b\|_2$&$\|x-x_{\textmd{org}}\|_2$&CPU time\\\hline
\multirow{4}{*}{$k=0.05n$}&1280&4.02e-04&	1.17e-05&	1.40 \\
&5120&	1.70e-03&	1.09e-05&	13.71 \\
&7680&	2.10e-03&	8.70e-06&	29.45\\
&~~10240~~&	3.40e-03&	1.09e-05&	52.30\\\hline
\multirow{4}{*}{$k=0.01n$}&1280&   1.94e-04& 5.52e-06& 0.555\\
&5120&   7.99e-04& 5.86e-06& 8.634\\
&7680&   1.36e-03& 6.85e-06& 18.80\\
&10240&  1.34e-03& 5.03e-06& 34.60\\\hline
\end{tabular}\par}}
\end{table}
\end{center}
%
\subsection{Computational Results: Recovery with Noise}
We now consider the recovery with noise: $$y=\Phi x+\xi,$$
where the noise $\xi$ obeys the normal distribution with zero expectation and $\sigma^2$ variance, namely $\xi\sim N(0,\sigma^2)$. Here we take $\sigma=0.01$. Under the noise case, from Figs \ref{figs11} and \ref{figs22}, one can not be difficult to see the comments below.
 \begin{itemize}
   \item For any $q$, sparsity derived from MRIL1 is closer to the true sparsity than that from RIL1 and IL1.
   \item When $q=0.1,0.25,0.5,0.6$, the results from RIL1 are excessively sparse, and then $q=0.75,0.9$, RIL1 begins to perform as well as the MRIL1 which always performs steadily well. However IL1 always does not obtain the true sparsity.
 \end{itemize}

 \makeatletter
          \def\@captype{figure}
          \makeatother
{\flushleft\includegraphics[width=9.7cm]{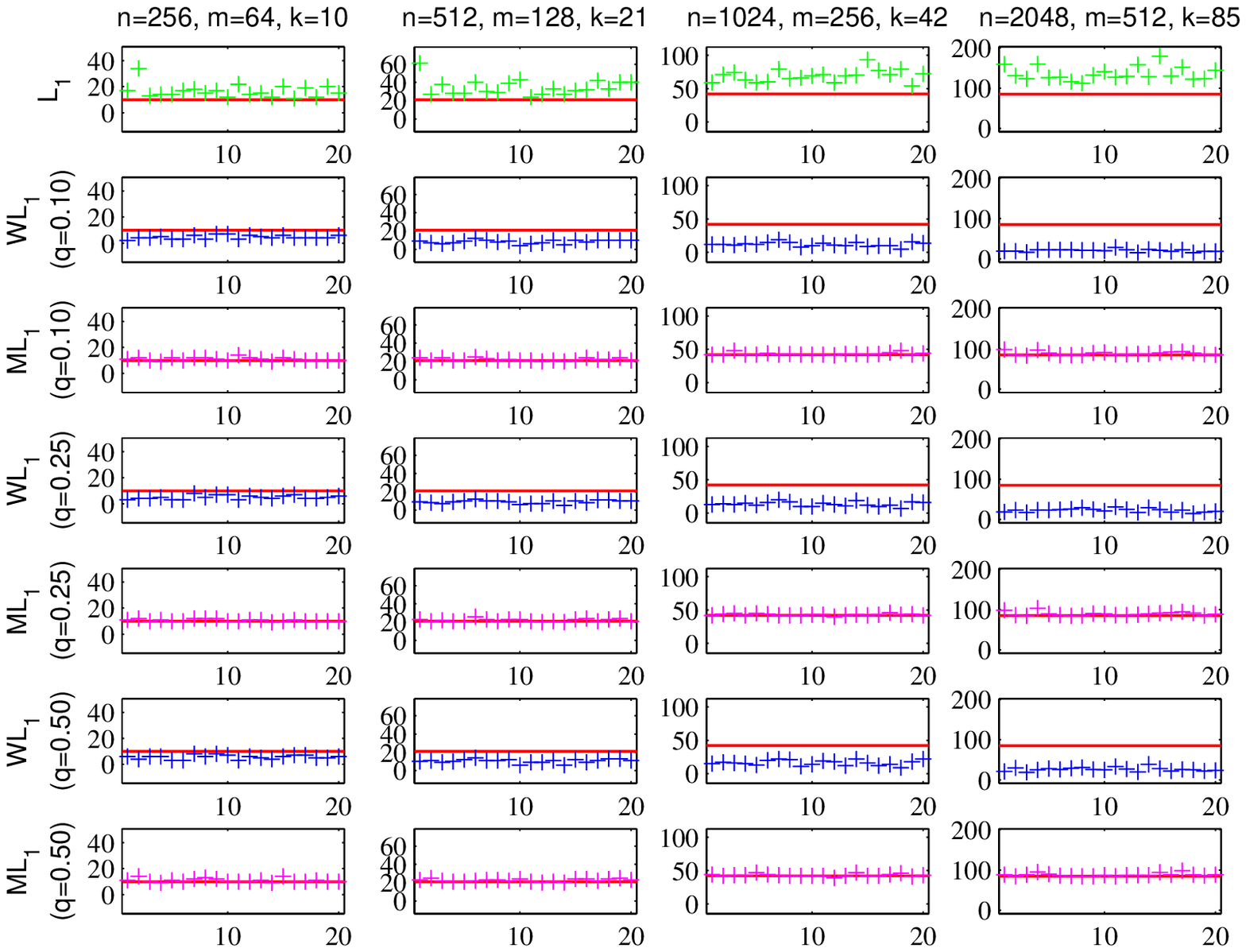}}
          \vspace{-8mm}
          \caption{Sparsity yielded by IL1, IRL1 and MIRL1 when $q=0.1, 0.25, 0.5$.\label{figs11}}

   \makeatletter
         \def\@captype{figure}
          \makeatother
          \begin{flushleft}
          \includegraphics[width=9.7cm]{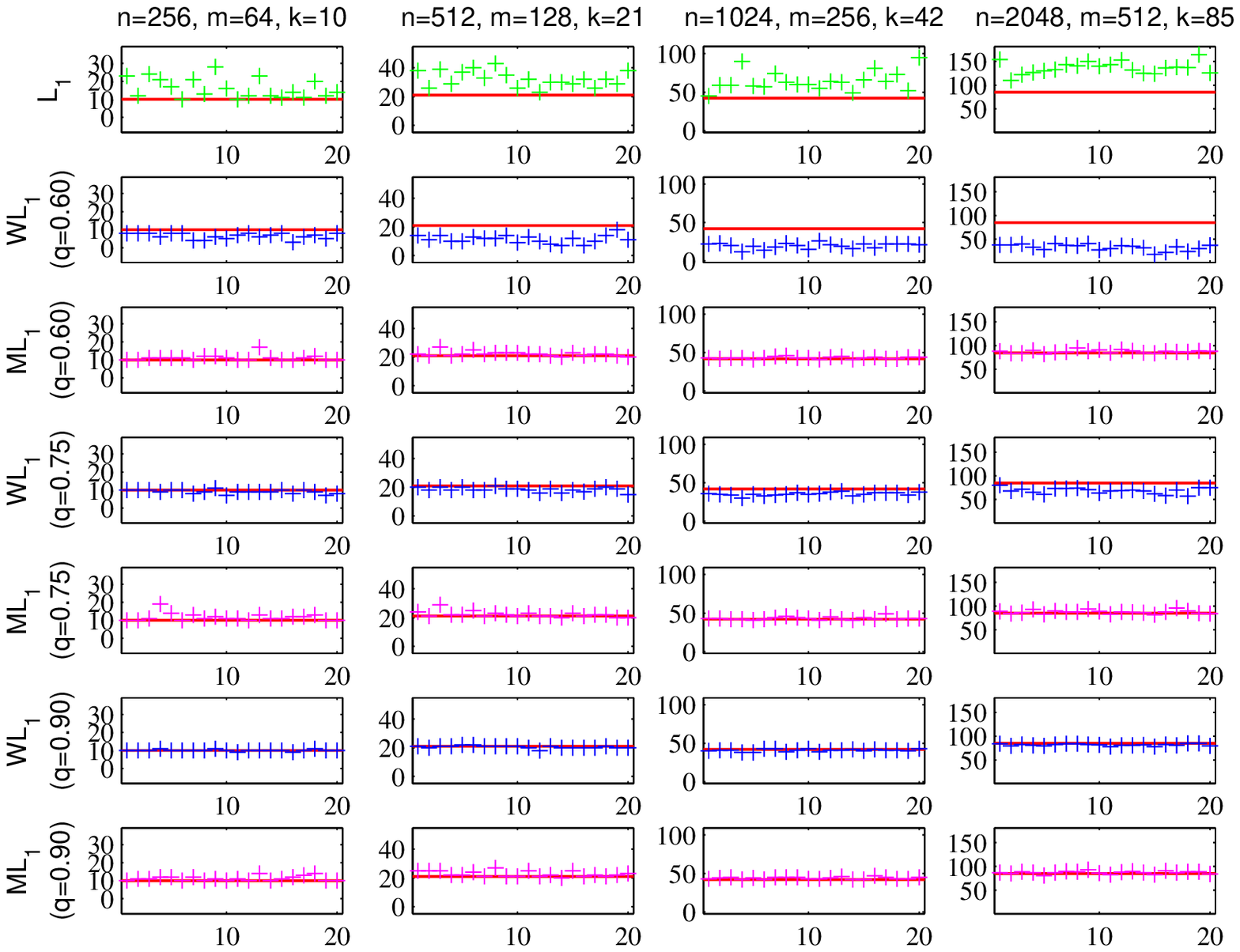}
          \end{flushleft}
         \vspace{-8mm}
          \caption{Sparsity yielded by  IL1, IRL1 and MIRL1 when $q=0.6, 0.75, 0.9$.\label{figs22}}
\vspace{5mm}

From Figs \ref{fige11}--\ref{figt22}, several comments can be derived.
 \begin{itemize}
   \item For any $q$, the average errors $\|\Phi x-b\|_2$ (or $\|x-x_{\textmd{orig}}\|_\infty$) by MRIL1 are  smaller than those of IL1 and  RIL1; Particularly, when $q=0.9$,  IRL1, MRIL1 and IL1 basically proceed identically well, which indicates RIL1 method is overly dependent on the parameter $q$;
    \item The average CPU time generated by RIL1 are smallest, and close behind is the MRIL1 for any $q$. IL1 costs the longest time to pursue the optimal solutions.
 \end{itemize}
 \makeatletter
          \def\@captype{figure}
          \makeatother
          \begin{center}
          \includegraphics[width=9.1cm]{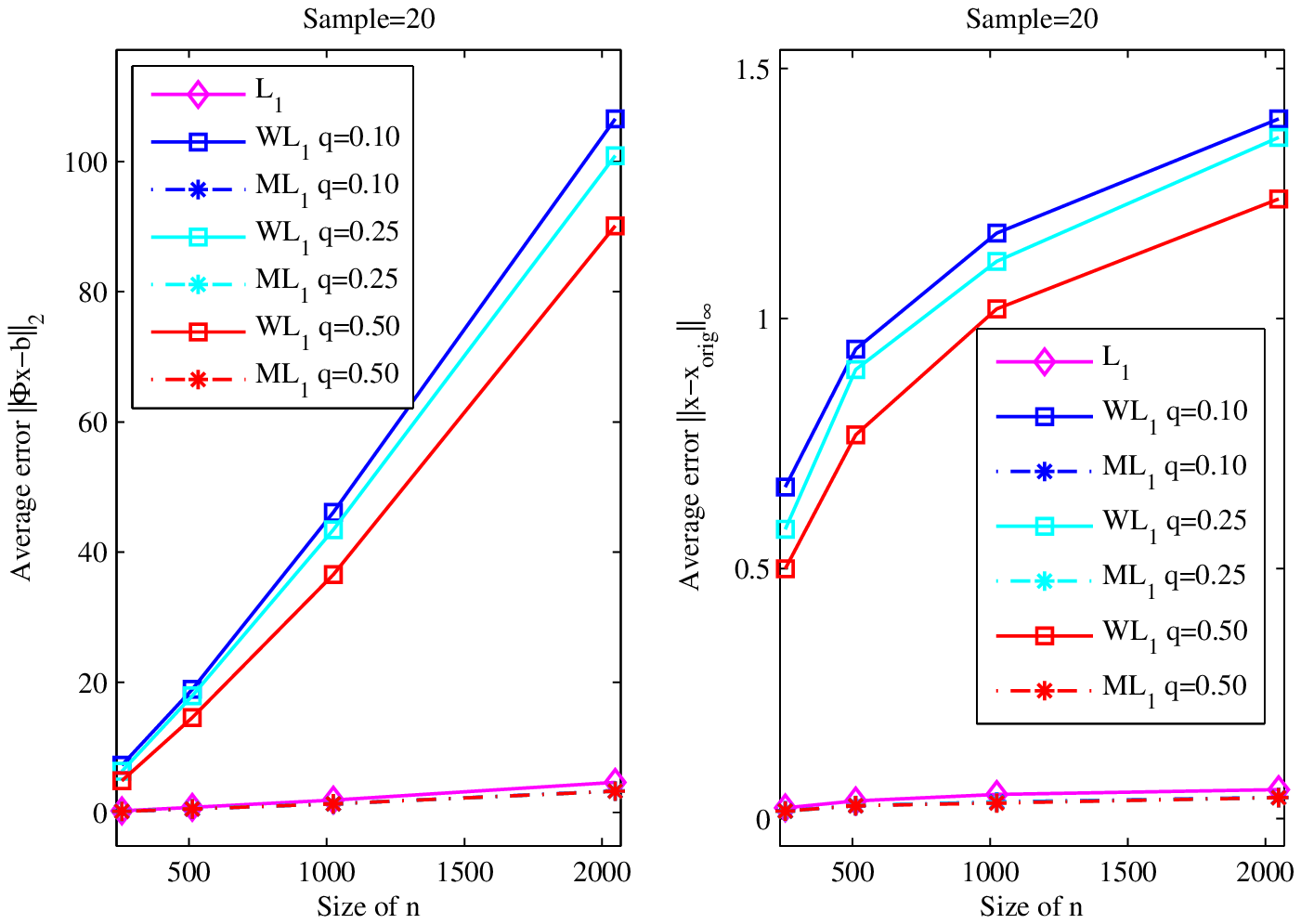}
          \vspace{-5mm}
          \caption{Error yielded by IL1, IRL1 and MIRL1 when $q=0.1, 0.25, 0.5$ .\label{fige11}}
        \end{center}
         \makeatletter
          \def\@captype{figure}
          \makeatother
          \begin{center}
          \includegraphics[width=9.1cm]{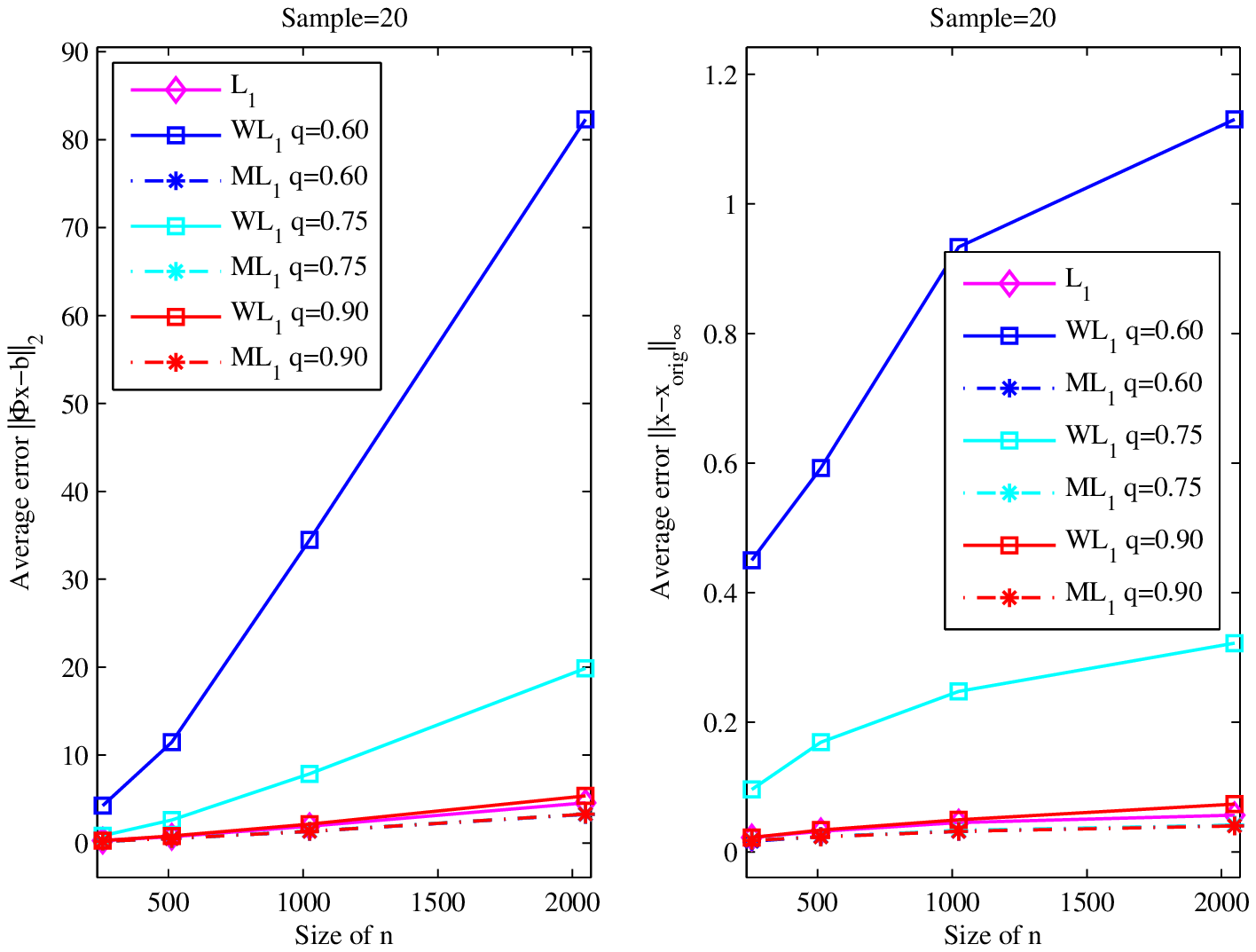}
          \vspace{-5mm}
          \caption{Error yielded by IL1, IRL1 and MIRL1 when $q=0.6, 0.75, 0.9$.\label{fige22}}
        \end{center}
         \makeatletter
          \def\@captype{figure}
          \makeatother
          \begin{center}
          \includegraphics[width=9.1cm]{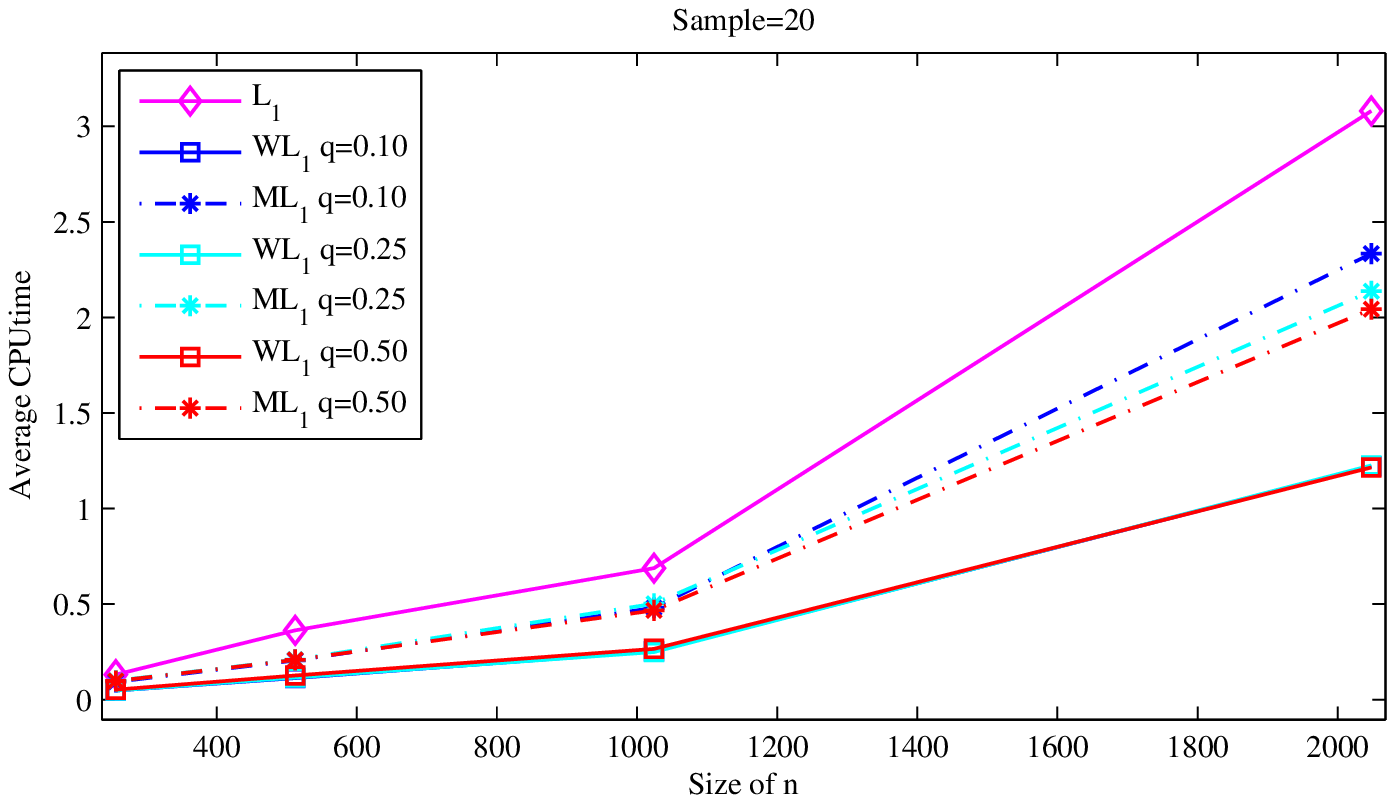}
          \vspace{-5mm}
          \caption{Time yielded by IL1, IRL1 and MIRL1 when $q=0.1, 0.25, 0.5$.\label{figt11}}
        \end{center}

         \makeatletter
          \def\@captype{figure}
          \makeatother
          \begin{center}
          \includegraphics[width=9.1cm]{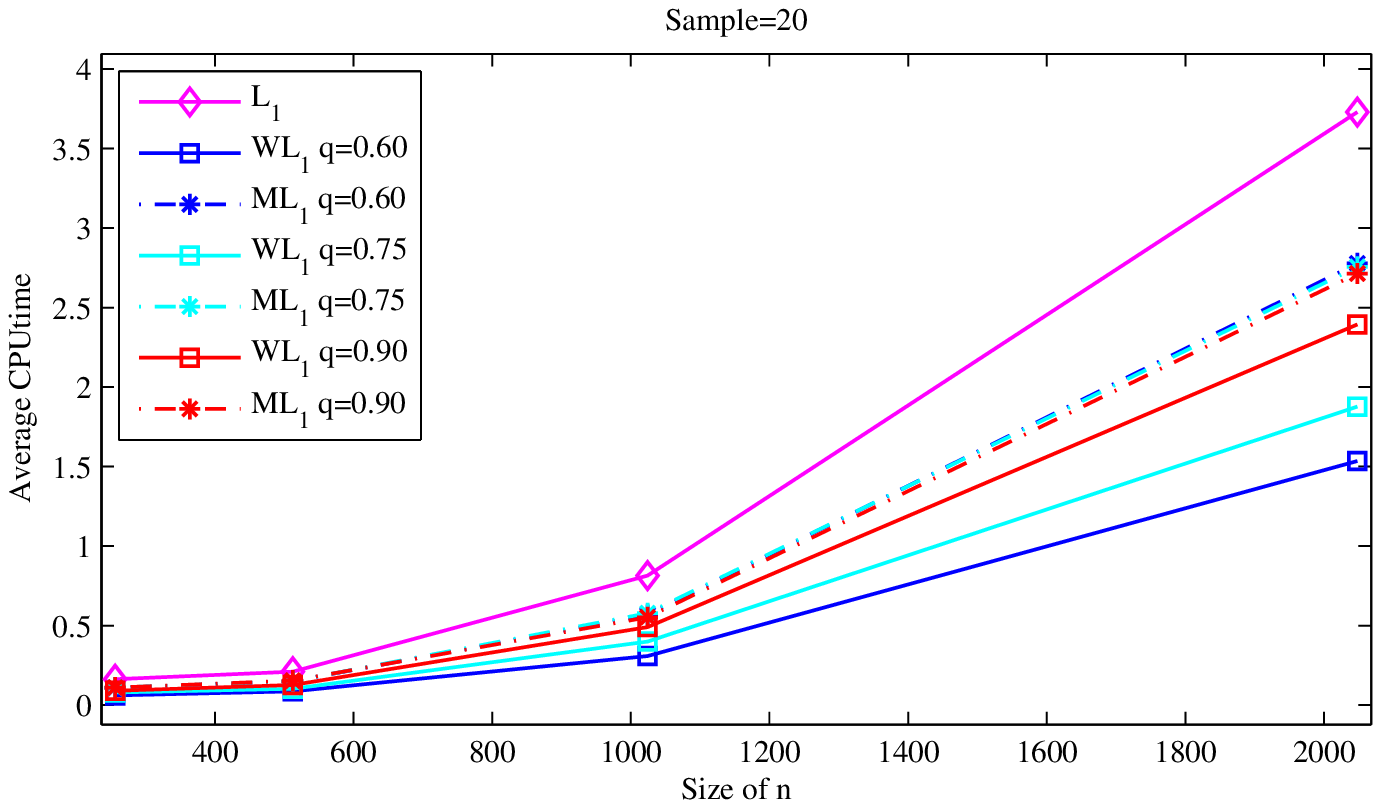}
          \vspace{-5mm}
          \caption{Time yielded by IL1, IRL1 and MIRL1 when $q=0.6, 0.75, 0.9$.\label{figt22}}
        \end{center}

\section{Conclusion}
In this  paper,  we have established weighted null space property  and RIC bounds through the weighted $\ell_1$ minimization for exact sparse recovery. The upper bounds on RICs from the weighted $\ell_1$ minimization are better in some cases than the current results from $\ell_1$ minimization, and moreover the way presented in this paper, in a sense, gives us a hint to construct the weight to pursue the sparse solution. As a consequence, these results strengthen the theoretical foundation of the reweighted $\ell_1$ minimization approach utilized extensively in sparse signal recovery. Moreover, the proposed method based on our new RIC theory provides an effective access to locate the none zero entries of the original sparse solution.
\section*{Acknowledgement}
The work was supported in part by the National Basic Research Program of China (2010CB732501), and the National Natural Science Foundation of China (11171018, 71271021).

\section{Appendix}

\noindent\textbf{Proof of Lemma \ref{lemma00}}\\
On one hand, it is obvious for Definition \ref{de1} to get  Definition \ref{de3}. On the other hand, if $\Phi$ satisfies the null space property defined by Definition \ref{de3}, that is, for all subsets $S\in\mathcal{C}_{n}^{k}$ it holds
\begin{eqnarray}\left\|  h_{S}\right\|_1<\left\| h_{S^C}\right\|_1\nonumber\end{eqnarray}
 for any $h\in \mathcal{N}_1$. For any $h'\in \mathcal{N}(\Phi)\setminus\{0\}$ with $\|h'\|_1=\varsigma>0$, we have $\|\frac{h'}{\varsigma}\|_1\in \mathcal{N}_1$ and thus
\begin{eqnarray}\left\| \left(h'/\varsigma\right)_{S}\right\|_1<\left\| \left(h'/\varsigma\right)_{S^C}\right\|_1\nonumber\end{eqnarray}
which is equal to
\begin{eqnarray}\left\| h'_{S}\right\|_1<\left\| h'_{S^C}\right\|_1.\nonumber\end{eqnarray}
Henceforth, $\Phi$ also satisfies the null space property defined by Definition \ref{de1}.\qed

\vspace{2mm}
\noindent\textbf{Proof of Theorem \ref{Theorem1}}\\
\noindent(\textbf{Sufficiency}) Let us assume the WNSP of order $k$ holds. For a given $\omega$, any $h\in \mathcal{N}(\Phi)$ with $\|h\|_1=\varsigma>0$ and all subsets $S\in\mathcal{C}_{n}^{k}$, from (\ref{wnsp}) it follows that $\left\|\omega\circ (h/\varsigma)_{S}\right\|_1<\left\|\omega\circ (h/\varsigma)_{S^C}\right\|_1$, which is equivalent to
\begin{eqnarray}\label{wnsp1}\left\|\omega\circ h_{S}\right\|_1<\left\|\omega\circ h_{S^C}\right\|_1.\end{eqnarray}
Hence, for any $k$-sparse vector $\hat{x}\in\mathbb{R}^n$, $h\in \mathcal{N}(\Phi)$ and all subsets $S\in\mathcal{C}_{n}^{k}$, from (\ref{wnsp1}) we obtain,
\begin{eqnarray}0&<&\sum_{i\in S^C}\omega_i|h_i|-\sum_{i\in S}\omega_i|h_i|\nonumber\\
&\leq&\sum_{i\in S^C}\omega_i|h_i|+\sum_{i\in S}\omega_i\left(|\hat{x}_i+h_i|-|\hat{x}_i|\right).\nonumber\end{eqnarray}
Since $\widehat{S}:=(\textmd{supp}(\hat{x})^{T},0)^{T}\in\mathcal{C}_{n}^{k}$,  together with the inequality above, we have
\begin{eqnarray}\|\omega\circ \hat{x}\|_1&=&\sum_{i\in \widehat{S}}\omega_i|\hat{x}_i|\nonumber\\
&<&\sum_{i\in \widehat{S}^C}\omega_i|h_i|+\sum_{i\in \widehat{S}}\omega_i|\hat{x}_i+h_i|\nonumber\\
&=&\|\omega\circ (\hat{x}+h)\|_1.\nonumber\end{eqnarray}
This established the required minimality of $\|\omega\circ x\|_{1}$.

(\textbf{Necessity}) Assume every $k$-sparse vector $\hat{x}\in\mathbb{R}^n$ is the unique solution of $\|\omega\circ x\|_{1}$ subject to $\Phi x=\Phi\hat{x}$. Then, in particular, for any $h\in \mathcal{N}_1$ and all subsets $S\in\mathcal{C}_{n}^{k}$, the $k$-sparse vector $h_S$  is the unique solution of $\|\omega\circ x\|_{1}$ subject to $\Phi x=\Phi h_S$. Since $\Phi h=0$, we have $\Phi h_{S}=\Phi (-h_{S^{C}})$, which means that
\begin{eqnarray}\label{ssc}\left\|\omega\circ h_{S}\right\|_{1}<\left\|\omega\circ h_{S^{C}}\right\|_{1}.\end{eqnarray}
The whole proof is completed immediately. \qed

\vspace{2mm}
\noindent\textbf{Proof of Lemma \ref{lemma1}}\\
If  $T_0$ defined as $(\ref{t0})$ uniquely exists, by denoting $T_1\in\mathcal{C}_{n}^{k}\setminus\{T_0\}$ as  \begin{eqnarray}\label{t1}(T_1,\widetilde{h}):=\underset{T\in\mathcal{C}_{n}^{k}\setminus\{T_0\},h\in\mathcal{N}_1}{\textmd{argmax}}\| h_{T}\|_1,\nonumber\end{eqnarray}
and taking $0<\frac{\left\|\widetilde{h}_{T_1}\right\|_1}{\left\|\widehat{h}_{T_0}\right\|_1}<\gamma<1$ by the uniqueness of $T_0$,
$$\|\omega\circ \widehat{h}_{T_0}\|_1=\gamma\| \widehat{h}_{T_0}\|_1>\|\widetilde{h}_{T_1}
\|_1\geq\left\|h_{T}\right\|_1\geq\left\|\omega\circ h_{T}\right\|_1$$
holds for any $h\in \mathcal{N}_1$ and any $T\in\mathcal{C}_{n}^{k}\setminus\{T_0\}$.

\noindent If $T_0$ exists but not uniquely, it is evident that $\omega$ defined as $(\ref{wt0})$ with $\gamma=1$  satisfies $(\ref{maxw})$.\qed

\vspace{2mm}
\noindent\textbf{Proof of Lemma \ref{lemma2}}\\
First we show the following fact based on our notations
\begin{eqnarray}\label{wh} \|\omega\circ \widehat{h}\|_1=\underset{h\in\mathcal{N}_1}{\min}\| \omega\circ h\|_1.\end{eqnarray}
Since $(\ref{wt0})$ with $0<\gamma\leq1$, for any $h\in \mathcal{N}_1$ we have
\begin{eqnarray}\|\omega\circ \widehat{h}\|_1&=&\|\omega\circ \widehat{h}_{T_0}\|_1+\|\widehat{h}_{T_0^{C}}\|_1\nonumber\\
&=&\gamma\| \widehat{h}_{T_0}\|_1+1-\| \widehat{h}_{T_0}\|_1\nonumber\\
&\leq&(\gamma-1)\| h_{T_0}\|_1+1\nonumber\\
&=&(\gamma-1)\| h_{T_0}\|_1+\| h_{T_0}\|_1+\| h_{T_0^{C}}\|_1\nonumber\\
&=&\|\omega\circ h_{T_0}\|_1+\|h_{T_0^{C}}\|_1\nonumber\\
&=&\|\omega\circ h\|_1,\nonumber\end{eqnarray}
where the first inequality is resulted from (\ref{t0}).

Then to prove WNSP, namely to show \begin{eqnarray} \left\|\omega\circ h_{T^C}\right\|_1>\left\|\omega\circ h_{T}\right\|_1\nonumber\end{eqnarray}
holds for any $h\in \mathcal{N}_1$ and any $T\in\mathcal{C}_{n}^{k}$. By the definition of $T_0$ in (\ref{t0}), if (\ref{t0nsp}) holds, that is
\begin{eqnarray}\| \omega\circ\widehat{ h}_{T_0^C}\|_1=\|\widehat{h}_{T_0^C}\|_1> \gamma\| \widehat{h}_{T_0}\|_1=\|\omega\circ \widehat{h}_{T_0}\|_1,\nonumber\end{eqnarray}
then for any $h\in \mathcal{N}_1$ and any $T\in\mathcal{C}_{n}^{k}$, we have
 \begin{eqnarray} \left\|\omega\circ h_{T^C}\right\|_1&=&\|\omega\circ h\|_1-\left\|\omega\circ h_{T}\right\|_1\nonumber\\
&\geq& \|\omega\circ \widehat{h}\|_1-\left\|\omega\circ h_{T}\right\|_1\nonumber\\
&=&\|\omega\circ \widehat{h}_{T_0}\|_1+\| \omega\circ\widehat{h}_{T_0^C}\|_1-\left\|\omega\circ h_{T}\right\|_1\nonumber\\
&>&2\|\omega\circ \widehat{h}_{T_0}\|_1-\left\|\omega\circ h_{T}\right\|_1\nonumber\\
&>&\|\omega\circ h_{T}\|_1,\nonumber\end{eqnarray}
the first and last inequalities follow from (\ref{wh}) and Lemma \ref{lemma1} respectively.\qed

\vspace{2mm}
\noindent\textbf{Proof of Theorem \ref{Theorem2}}\\
From Lemma \ref{lemma2} and  Theorem \ref{Theorem1}, to pursue WNSP, we only need to check $(\ref{t0nsp})$, that is
\begin{eqnarray}\|\widehat{h}_{T_0^C}\|_1>\gamma\| \widehat{h}_{T_0}\|_1.\nonumber\end{eqnarray}
For simplicity we shortly denote hereafter $h=\widehat{h}$, from above inequality we suppose on the contrary that
\begin{eqnarray} \|h_{T_0^C}\|_1\leq\gamma\left\|h_{T_0}\right\|_1.\nonumber\end{eqnarray}
 By setting $\beta:=\left\|h_{T_0}\right\|_1/k$, then we have
 $$\|h_{T_0^C}\|_1\leq \gamma k\beta.$$
We now divide $h_{T_0^C}$ into two parts, $h_{T_0^C}= h^{(1)} + h^{(2)}$, where
\begin{numcases}{h^{(1)}_i=}
(h_{T_0^C})_i,~~~|(h_{T_0^C})_i|>\beta/t,\nonumber\\
0,~~~~~~~~~~~\text{otherwise},\nonumber\end{numcases}
\begin{numcases}{h^{(2)}_i=}
(h_{T_0^C})_i,~~~|(h_{T_0^C})_i|\leq\beta/t,\nonumber\\
0,~~~~~~~~~~~\text{otherwise},\nonumber\end{numcases}
and $t>0$ satisfies $\gamma kt$ being an integer. Therefore $h^{(1)}$ is $\gamma kt$-sparse as a result of facts that $\|h^{(1)}\|_1\leq\|h_{T_0^C}\|_1\leq \gamma k\beta$ and all non-zero entries of $h^{(1)}$ has magnitude larger than $\frac{\beta}{t}$. By letting $\|h^{(1)}\|_0 = m$, then it produces
\begin{eqnarray}&&\|h^{(2)}\|_1=\|h_{\overline{}T_0^C}\|_1-\|h^{(1)}\|_1\leq\left[\gamma kt-m\right]\beta/t,~~~~~~\\
&&\|h^{(2)}\|_\infty\leq\beta/t.\end{eqnarray}
Applying Lemma \ref{lemma0} with $s=\gamma kt-m$, it makes $h^{(2)}$ be expressed as a convex combination of sparse vectors, i.e., $$h^{(2)} =\sum_{i=1}^{N}\lambda_iu_i,$$ where $u_{i}$ is $(\gamma kt-m)$-sparse, $\|u_{i}\|_1=\|h^{(2)}\|_1, \|u_{i}\|_\infty\leq\beta/t,~i=1,2,\cdots,N$. Henceforth,
\begin{eqnarray}\label{12}\|u_{i}\|_2^2&\leq&(\gamma kt-m)\|u_{i}\|_\infty^2\leq\frac{\gamma k}{t}\beta^{2}\nonumber\\
&\leq&\frac{\gamma }{t}\|h_{T_0}\|_2^2\leq\frac{\gamma}{t}\|h_{T_0}+h^{(1)}\|_2^2,\end{eqnarray}
where the third and last inequalities are as the consequences of the $\|h_{T_0}\|_1\leq\sqrt{k}\|h_{T_0}\|_2$, and disjoint supports of $h_{T_0}$ and $h^{(1)}$ respectively.

For any $\mu\geq0$, denoting $\eta_i = h_{T_0}+ h^{(1)}+\mu u_{i}$, we obtain
\begin{eqnarray}
&&\sum_{j=1}^{N}\lambda_j\eta_j-\eta_i/2\nonumber\\
&=&h_{T_0}+ h^{(1)}+\mu h^{(2)}-\eta_i/2\nonumber\\
\label{a}&=&(\frac{1}{2}-\mu)\left(h_{T_0}+ h^{(1)}\right)-\mu u_i/2+\mu h,\end{eqnarray}
where $\eta_i,\sum_{i=j}^{N}\lambda_j\eta_j-\frac{1}{2}\eta_i-\mu h$ are all $\left(\gamma kt+k\right)$-sparse vectors thanks to the sparsity of $\|h_{T_0}\|_0\leq k$,~$\|h^{(1)}\|_0=m$ and $\|u_{i}\|_0\leq \gamma kt-m$.
Since $\Phi h = 0$, together with (\ref{a}), we have $$\Phi(\sum_{j=1}^{N}\lambda_j\eta_j-\frac{1}{2}\eta_i)=\Phi((\frac{1}{2}-\mu)(h_{T_0}+ h^{(1)})-\frac{1}{2}\mu u_i).$$
Following the proof of Theorem 1.1 in \cite{CZ5}, we easily elicit
\begin{eqnarray}\label{ab}
\sum_{i=1}^{N}\lambda_i\|\Phi(\sum_{j=1}^{N}\lambda_j\eta_j-\frac{1}{2}\eta_i)\|_2^2=\frac{1}{4}\sum_{i=1}^{N}\lambda_i\|\Phi\eta_i\|_2^2.
\end{eqnarray}
Setting $\mu=\frac{\sqrt{(t+\gamma)t}-t}{\gamma}>0$, if it holds that
\begin{eqnarray}\label{aa}\delta:=\delta_{\gamma kt+k}<\sqrt{\frac{t}{t+\gamma}},\end{eqnarray}
then combining (\ref{ab}) with (\ref{aa}), we get
\begin{eqnarray}
0&=&\sum_{i=1}^{N}\lambda_i\|\Phi((\frac{1}{2}-\mu)(h_{T_0}+h^{(1)})-\frac{1}{2}\mu u_i)\|_2^2\nonumber\\
&&-\frac{1}{4}\sum_{i=1}^{N}\lambda_i\|\Phi\eta_i\|_2^2\nonumber\\
&\leq&(1+\delta)\sum_{i=1}^{N}\lambda_i[(\frac{1}{2}-\mu)^2\|h_{T_0}+h^{(1)}\|_2^2+\frac{\mu^{2}}{4}\|u_i\|_2^2]\nonumber\\
&&-\frac{1-\delta}{4}\sum_{i=1}^{N}\lambda_i(\|h_{T_0}+ h^{(1)}\|_2^2+\mu^{2}\|u_i\|_2^2)\nonumber\\
&=&\sum_{i=1}^{N}\lambda_i[((1+\delta)(\frac{1}{2}-\mu)^2-\frac{1-\delta}{4})\cdot\nonumber\\
&&\|h_{T_0}+h^{(1)}\|_2^2+\frac{1}{2}\delta\mu^{2}\|u_i\|_2^2]\nonumber\end{eqnarray}
\begin{eqnarray}&\leq&\sum_{i=1}^{N}\lambda_i\|h_{T_0}+h^{(1)}\|_2^2\cdot\nonumber\\
\label{c}&&\left[\mu^2-\mu+\delta(\frac{1}{2}-\mu+(1+\frac{\gamma}{2t})\mu^{2})\right]\nonumber\\
&=&\|h_{T_0}+h^{(1)}\|_2^2\cdot\nonumber\\
&&\left[\mu^2-\mu+\delta(\frac{1}{2}-\mu+(1+\frac{\gamma}{2t})\mu^{2})\right]\nonumber\\
&=&\|h_{T_0}+h^{(1)}\|_2^2\left(\frac{1}{2}-\mu+(1+\frac{\gamma}{2t})\mu^{2}\right)\cdot\nonumber\\
&&\left[\delta-\sqrt{\frac{t}{t+\gamma}}\right]\nonumber\\
&<&0, \nonumber\end{eqnarray}
where the second inequality is derived from (\ref{12}). Obviously, this is a contradiction.

 For condition (\ref{aa}), setting $t=\frac{a-1}{\gamma}$, it follows that
 \begin{eqnarray}\delta_{ak}< \sqrt{\frac{a-1}{a-1+\gamma^2}}.\nonumber\end{eqnarray}
Hence we complete the proof. \qed

\vspace{3mm}
In order to prove the result Theorem \ref{Theorem2}, we need another important concept in the RIP framework, the restricted orthogonal constants (ROC), proposed in \cite{CT1}.
\begin{Definition} Define the restricted orthogonal constants (ROCs) of order $k_1, k_2$ for a matrix $\Phi\in\mathbb{R}^{m\times n}$ as the smallest non-negative number $\theta_{k_1, k_2}$ such that
\begin{eqnarray}\label{theta} \left|\left\langle\Phi h_1,\Phi h_2\right\rangle\right|\leq\theta_{k_1, k_2}\|h_1\|_2\|h_2\|_2\end{eqnarray}
for all $k_1$-sparse vector $h_1\in\mathbb{R}^{n}$ and $k_2$-sparse vector $h_2\in\mathbb{R}^{n}$ with disjoint supports.\end{Definition}

The next lemma blending with Lemmas 3.1, 5.1 and 5.4 in \cite{CZ4} centers on several properties of the restricted orthogonal constants (ROCs).
\begin{Lemma}\label{lemma5} Let $k_1, k_2, k \leq n, \lambda\geq0$ and $\mu\geq1$ such that $\mu k_2$ is an integer. Suppose $h_1,h_2\in\mathbb{R}^{n}$ have disjoint supports and $h_1$ is
$k_1$-sparse. If $\|h_2\|_1\leq\lambda k_2$ and $\|h_2\|_\infty\leq\lambda$, then the restricted orthogonal constants satisfy
\begin{eqnarray}\label{p1}&& \left|\left\langle\Phi h_1,\Phi h_2\right\rangle\right|\leq\theta_{k_1, k_2}\|h_1\|_2\lambda\sqrt{k_2},\\
\label{p2} &&\theta_{k_1, \mu k_2}\leq\sqrt{\mu}\theta_{k_1, k_2},\end{eqnarray}
and
\begin{numcases}{\theta_{k,k}<}
\label{p3}~~~~2\delta_{k},  ~~~~~~~~\text{for any even}~ k\geq2,\\
\label{p4}\frac{2k}{\sqrt{k^2-1}}\delta_{k},~~~~\text{for any odd}~k\geq3.\end{numcases}\end{Lemma}

\vspace{2mm}
\noindent\textbf{Proof of Theorem \ref{Theorem3}}\\
Similar to the proof of Theorem \ref{Theorem1}, we suppose on the contrary that $\widehat{h} \in \mathcal{N}_1$ (also shortly denote hereafter $h=\widehat{h}$) such that
$$\|h_{T_0^C}\|_1\leq \gamma\|h_{T_0}\|_1.$$
 Setting $\beta=k^{-1}\|h_{T_0}\|_{1}$, then we have $\|h_{T_0^C}\|_1\leq \gamma k \beta\leq\lceil \gamma k\rceil \beta$ and $\|h_{T_0^C}\|_\infty\leq \beta$. In fact, if $\|h_{T_0^C}\|_\infty>\beta$, then (\ref{t0}) will contribute to $k\beta=\|h_{T_0}\|_1\geq k\|h_{T_0^C}\|_\infty>k\beta$. Thus it follows that
\begin{eqnarray}
|\langle\Phi h_{T_0},\Phi h_{T_0^C}\rangle|&\leq&\theta_{k,\lceil\gamma k\rceil}\|h_{T_0}\|_2\sqrt{\lceil \gamma k\rceil} \beta\nonumber\\
&\leq&\theta_{k,k}\sqrt{\frac{\lceil \gamma k\rceil }{k}}\|h_{T_0}\|_2\sqrt{\lceil \gamma k\rceil} \beta\nonumber\\
&\leq&\theta_{k,k}\frac{\lceil \gamma k\rceil }{k}\|h_{T_0}\|_2^2,\nonumber\end{eqnarray}
where the first and second inequalities hold by ($\ref{p1}$) and ($\ref{p2}$) in Lemma \ref{lemma5} respectively, and the last inequality is derived from H$\ddot{o}$lder inequality, i.e., $\|h_{T_0}\|_1\leq\sqrt{k}\|h_{T_0}\|_2$. If it holds
\begin{eqnarray}\label{cond2}\delta_{k}+\theta_{k,k}\frac{\lceil \gamma k\rceil }{k}<1,\end{eqnarray}
 we have
\begin{eqnarray}
0&=&\left|\left\langle\Phi h_{T_0},\Phi h\right\rangle\right|\nonumber\\
&\geq&\left|\left\langle\Phi h_{T_0},\Phi h_{T_0}\right\rangle\right|-|\langle\Phi h_{T_0},\Phi h_{T_0^C}\rangle|\nonumber\\
&\geq&(1-\delta_{k})\|h_{T_0}\|_2^2-\theta_{k,k}\frac{\lceil \gamma k\rceil }{k}\|h_{T_0}\|_2^2\nonumber\\
&=&\left(1-\delta_{k}-\theta_{k,k}\frac{\lceil \gamma k\rceil }{k}\right)\|h_{T_0}\|_2^2\nonumber\\
&>&0.\nonumber\end{eqnarray}
Obviously, this is a contradiction. By (\ref{p3}) and (\ref{p4}) in Lemma \ref{lemma5}, when $k\geq2$ is even, it yields
$$\delta_{k}+\theta_{k,k}\frac{\lceil \gamma k\rceil}{k}<\left(1+\frac{2\lceil \gamma k\rceil}{k}\right)\delta_{k},$$
and when $k\geq3$ is odd, it generates that
$$\delta_{k}+\theta_{k,k}\frac{\lceil \gamma k\rceil}{k}<\left(1+\frac{2\lceil \gamma k\rceil}{\sqrt{k^2-1}}\right)\delta_{k}.$$
Therefore the theorem is accomplished thanks to conditions (\ref{even}) and (\ref{odd}) enabling (\ref{cond2}) to hold. \qed

\vskip2mm
\noindent\footnotesize{\textbf{Shenglong Zhou} is a PhD student in Department of Applied Mathematics, Beijing Jiaotong University.  He received his BS  degree from Beijing Jiaotong University of information and computing science in 2011. His research field is theory and methods for optimization.}
\vskip2mm
\noindent\footnotesize{\textbf{Naihua Xiu}} is a Professor in Department of Applied Mathematics, Beijing Jiaotong University.  He received his  PhD  degree in Operations Research from Academy Mathematics and System Science of the Chinese Academy of Science in 1997. He was a Research Fellow of City University of Hong Kong from 2000 to 2002, and he was a Visiting Scholar of University of Waterloo from 2006 to 2007. His research interest includes variational analysis, mathematical optimization,  mathematics of operations research.
\vskip2mm
\noindent\footnotesize{\textbf{Yingnan Wang} is a research assistant in Department of Applied Mathematics, Beijing
Jiaotong University. She received her PhD degree in Operations
Research from Beijing Jiaotong University in 2011. From 2011 to 2013, she
was a Post-Doctoral Fellow of Department of Combinatorics and Optimization,
Faculty of Mathematics, University of Waterloo, Canada. Her research
interests are in sparse optimization, non-smooth optimization and analysis, robust optimization.}
\vskip2mm
\noindent\footnotesize{\textbf{Lingchen Kong}} is an associate Professor in Department of Applied
Mathematics, Beijing Jiaotong University.  He received his PhD  degree in Operations Research from Beijing Jiaotong University in 2007.
From 2007 to 2009,  he was a Post-Doctoral Fellow of Department of
Combinatorics and Optimization, Faculty of Mathematics, University
of Waterloo, Canada. His research interests are in sparse optimization, mathematics of operations
research.

\end{document}